\documentclass[twocolumn,showpacs,prb,amsmath,amssymb]{revtex4}
\usepackage{graphicx}% Include figure files
\usepackage{dcolumn}% Align table columns on decimal point
\usepackage{bm}% bold math

\begin{document}
\draft
%\twocolumn[\hsize\textwidth\columnwidth\hsize\csname @twocolumnfalse\endcsname

\title{
Theoretical study of 
a localized quantum spin reversal by the sequential injection of spins 
in a spin quantum dot
%sequential injection of electrons 
%Sequentially injected electrons
 }
\author{
Satoshi~Kokado$^{1}$
\footnote{Electronic mail: tskokad@ipc.shizuoka.ac.jp}, 
Kazumasa~Ueda$^1$, 
Kikuo~Harigaya$^2$, and 
Akimasa~Sakuma$^3$
}
\address{
$^1$Faculty of Engineering, Shizuoka University, Hamamatsu 432-8561, Japan \\
$^2$Nanotechnology Research Institute, AIST, Tsukuba 305-8568, Japan \\
$^3$Graduate School of Engineering, Tohoku University, Sendai 980-8579, Japan
}
\date{\today}

%\maketitle

\begin{abstract}
This is a theoretical study of the reversal of 
a localized quantum spin induced 
by sequential injection of spins 
for a spin quantum dot that has a quantum spin. 
%having the quantum spin. 
%where the dot has a localized quantum spin. 
%A system to investigate the reversal is 
The system consists of 
%to investigate the reversal is 
``electrode/quantum well(QW)/dot/QW/electrode" junctions, 
in which the left QW has an energy level 
of conduction electrons with only up-spin. 
%the electrodes are nonmagnetic. 
%it is connected to the left and right quantum wells which are contacted with the respective electrodes. 
We consider a situation in which 
up-spin electrons are sequentially injected 
from the left electrode into the dot through the QW 
and an exchange interaction acts between 
the electrons and the localized spin. 
To describe the sequentially injected electrons, 
we propose a simple method 
based on approximate solutions 
from the time-dependent Schr$\ddot{\rm o}$dinger equation. 
Using this method, 
it is shown that 
the spin reversal occurs 
when the right QW has energy levels 
of conduction electrons with only down-spin. 
%It should be emphasized that 
In particular, 
the expression of the reversal time of a localized spin is derived 
and the upper and lower limits of the time are clearly expressed. 
This expression is expected 
%believed 
%appears 
to be useful 
for a rough estimation of the minimum relaxation time 
of the localized spin to achieve the reversal. 
%We also obtain characteristic time dependences of 
We also obtain analytic expressions for 
%characteristic time dependences of 
the expectation value of the localized spin and 
%and 
%the time dependences of 
%the probability density and 
the electrical current 
%, and 
%the number of electrons in the right QW and the right electrode, 
as a function of time. 
%when the localized spin possesses $S \ge 1$ and the right electrode interface has energy levels of conduction electrons with only down-spin. 
%using a simple approach 
%based on the time dependent Schr$\ddot{\rm o}$dinger equation. 
%It is finally shown that 
In addition, we found that 
a system with the non-magnetic right QW 
exhibits spin reversal or non-reversal 
depending on the exchange interaction. 
\end{abstract}
%\pacs{72.25.Ba} 
%%%%%%%%%%%%%%%%%vskip2pc] 

\pacs{75.60.Jk,73.63.Kv,85.75.-d}% PACS, the Physics and Astronomy

\maketitle

\section{Introduction}
\label{introduction}

Magnetization reversal by spin injection (MRSI), 
in which the magnetization of a ferromagnet (FM) is reversed 
by injecting a spin-polarized current into the FM,~\cite{Slonczewski, Berger, Myers, Zhang, Heide, Hayakawa, Suzuki, Inomata, Katine, Grollier} 
is one of the most interesting topics in the field of 
spin-dependent transport and spin electronics 
for the following reasons: 
%because 
First, the reversal is induced by the current 
and 
not by the conventional method, such as the application of magnetic fields. 
Second, such a phenomenon is expected 
to have potential applications for 
%the spin electronics devices. 
%in 
%For instance, we take 
the writing of data in the magnetic memory. 
%Note that the control of the magnetization without using external magnetic fields enables high density and low power consumption memories. 
%the writing of data in the magnetic random access memory. 
%which 
%, where 
%An element of this memory 
%whose elements consist of FM/nonmagnetic layer/FM junctions, where current flows perpendicular to the plane. 
%Note also that 
%The control of the magnetization without using external magnetic fields enables high density and low power consumption memories. 

Originally, 
the MRSI was theoretically predicted by 
Slonczewski \cite{Slonczewski} and Berger.~\cite{Berger} 
%theoretically predicted 
Their systems are FM/non-magnetic layer/FM junctions, 
in which the current flows perpendicularly to the plane. 
%In their theoretical models, 
Here, the magnetization of the FM is described 
by classical spins, 
and an exchange interaction acts between 
the magnetization 
%magnetic moment of the FM 
and the electron spin.~\cite{Slonczewski, Berger} 
When the magnetization of one FM is free to change its direction 
and that of the other FM is pinned, 
%consisting of these junctions. 
%In these junctions, 
% represents that 
%and 
%As the writing method, 
%In the writing process, 
%The magnetization of the latter FM (free layer) is reversed by injecting the spin-polarized current from the former FM (pinned layer). 
the magnetization of the former can be changed to 
become 
either parallel or antiparallel to that of the latter 
%the pinned layer, 
depending on the direction of the current.~\cite{Slonczewski} 
% to the right one. 
%In particular, 
%Furthermore, 
In addition, 
some theoretical studies, in which the magnetization of the FM 
is described by the classical spins, have been reported 
since then.~\cite{Heide, Zhang}

%As results, the Slonczewski's model \cite{Slonczewski} showed that the MRSI was realized by the exertion of a current-induced torque, and the Berger's model~\cite{Berger} presented that the MRSI was associated with the stimulated emission of spin waves in the vicinity of an interface between ferromagnetic and nonmagnetic layers. 

Quite recently, the MRSI has been experimentally observed 
in FM/non-magnetic layer/FM junctions~\cite{Myers, Hayakawa, Suzuki, Inomata} 
and pillars.~\cite{Katine, Grollier} 
%~\cite{Myers, Hayakawa, Suzuki, Inomata} and pillars~{Katine, Grollier} 
%\cite{Myers} 
%which is an element of the MRAM; 
%Co/Cu/Co,~\cite{Myers} 
%CoFeB/MgO/CoFeB,~\cite{Hayakawa} 
%FeNi/Cu/FeNi,~\cite{Otani} 
%Co$_{0.75}$Fe$_{0.25}$/Cu/Co$_{0.75}$Fe$_{0.25}$,~\cite{Suzuki} 
%Co$_{90}$Fe$_{10}$/Cu/Co$_{90}$Fe$_{10}$/Ru/Co$_{90}$Fe$_{10}$~\cite{Inomata} 
%(Ga,Mn)As/ AlAs/(Ga,Mn)As,~\cite{Munekata} 
%permalloy particle/Cu/permalloy,~\cite{Otani} 
%$L1_{0}$-FePt/Fe/Au/$L1_{0}$-FePt~\cite{Takanashi} 
%Co/Cu/Co pillars,~\cite{Katine, Grollier} 
%and so on. 
%Furthermore, 
The experimental results have been often analyzed 
%explained 
%by using the above models 
by using the theoretical model~\cite{Slonczewski} 
based on the classical spins. 
%The MRSI has been often analyzed by using theoretical models, 
% 
The magnetization of the FM appears to be fairly well described 
by the classical spins.

On the other hand, we expect that 
magnetic materials will undergo transformation from classical spin systems 
into quantum spin ones 
along with miniaturization towards high-density integration devices 
in the future. 
%As such quantum spin systems, 
%we give, 
For example, 
such quantum spin systems are 
Mn$_{12}$ magnetic molecules.~\cite{Mn12_1, Mn12_2, Mn12_3, Awaga, Caneschi, Sessoli} 
A Mn$_{12}$ molecule possesses an effective spin of $S$=10 
due to an antiferromagnetic interaction between 
%which is formed by 
the eight Mn$^{3+}$ ($S$=2) ions and the four Mn$^{2+}$ ($S$=3/2) ions,~\cite{Caneschi, Sessoli, Mn12_2} 
%%with an antiferromagnetic interaction between the both, 
and it 
%this molecule 
has a uniaxial anisotropy energy, 
$-|D|S_z^2$, with $D$ being an anisotropy constant 
with the magnitude of 0.7 K.~\cite{Caneschi, Sessoli} 
The anisotropy energy shows a bistable potential between 
$S_z$=10 and $-$10 states. 
%the up and down-spin states. 
%As a characteristic phenomenon due to this potential, 
Regarding a characteristic phenomenon due to this potential, 
the quantum tunneling of magnetization 
%due to this potential 
%between their spin states 
has been experimentally observed under the magnetic fields.~\cite{Mn12_2} 
This phenomenon has been analyzed 
by using the quantum spin model 
with anisotropy energy.~\cite{Mn12_3} 
%describing the bistabe potential. 
%Furthermore, have been 
%studied extensively both experimentally and theoretically. 

%By utilizing such quantum spin systems, 
%By introducing such quantum spin systems 
%When such quantum spin systems are introduced 
%to the junctions with an electrode of the spin injection, 
If the spin-polarized current can be injected into 
such quantum spin systems, 
% are introduced to the junctions with an electrode of the spin injection, 
the conventional magnetization reversal 
%by the spin injection 
%may be replaced 
may be replaced 
by the localized quantum spin reversal. 
%by spin injection 
%in the future. 
In the present condition, 
however, 
%it is the present condition that 
very few theoretical studies for quantum spin reversal 
have been reported. 
%An investigation using a simple model is desirable 
%thought to be so significant 
%as a demonstration. 
%In particular, 
Our primary focus is on 
what models bring about the quantum spin reversal 
and the length of the reversal time of the localized spin. 
The latter is important for the purpose of 
roughly estimating 
%roughly estimating 
the minimum relaxation time of the localized spin 
to achieve the reversal.

In this paper, we examined the quantum spin reversal 
induced by the sequential injection of spins for a spin quantum dot 
having the quantum spin. 
%it is connected to the left quantum well ($L$) and the right one ($R$) which are 
%contacted with the respective electrodes. 
%where the dot has a localized quantum spin. 
%A system to investigate the reversal is 
The system consists of 
``electrode/quantum well(QW)/dot/QW/electrode" junctions, 
where the left QW ($L$) has an energy level of conduction electrons 
with only up-spin. 
% and the electrodes are nonmagnetic. 
We considered a situation in which 
up-spin electrons were sequentially injected 
from the left electrode into the dot through the $L$. 
%through a spin polarized quantum well (QW). 
% , and an exchange interaction acts between the electrons and the localized spin. 
To describe the sequentially injected electrons, 
we first proposed a simple method 
based on the time-dependent Schr$\ddot{\rm o}$dinger equation. 
Using this method, 
we obtained 
%Further, 
an expression of the reversal time of the localized spin 
%and showed the upper and lower limits of this time, 
%is obtained 
when 
%the $L$ had an energy level of conduction electrons with up-spin and 
the right QW ($R$) had energy levels of conduction electrons 
with only down-spin. 
Furthermore, 
%the time dependence of 
analytic expressions for 
the expectation value of the localized spin and 
%the time dependences of 
%the probability density and 
the electrical current 
%and the number of electrons in the $R$ and the right electrode
were obtained as a function of time. 
%It was also found that 
We also found 
the spin reversal occurred for the case of the specific exchange integral 
even when the $R$ was non-magnetic.

%The plan of the paper is as follows. 
The present paper is organized as follows. 
In Sec. \ref{Model}, 
we present the model of the spin quantum dot 
and assumptions for the sequential injection of spins. 
%processes. 
% for calculations. 
%In Sec., we derive expressions for 
In Sec. \ref{theo_des}, 
we provide a theoretical formulation; 
%derive expressions for 
%In particular, 
the wave function, 
%the number of electrons in the right electrode, 
the electrical current, and 
the expectation value of localized spin are derived. 
In Sec. \ref{appl1}, 
this theory is applied to the case of spin-polarized $R$, 
%having the energy levels of conduction electrons with only down-spin, 
%with only down-spin levels, 
while Sec. \ref{appl2} is the application to the case of non-magnetic $R$. 
In Sec. \ref{comments}, we make a proposal 
for a model with reversible switching and discuss the experimental aspects. 
Section \ref{conclusion} is the conclusion, 
and Sec. \ref{Appendix} is the appendix, 
which includes information about the bias and gate voltages.

\section{Model}
\label{Model}

\subsection{Spin quantum dot}

\begin{figure}[ht]
\begin{center}
\includegraphics[width=0.75\linewidth]{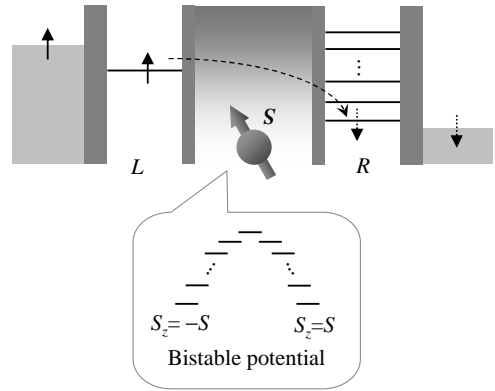}
\caption{
%The spin quantum dot. 
Schematic illustration of 
``electrode/$L$/spin quantum dot/$R$/electrode" junctions. 
%Here, 
%Here, $L$ ($R$) represents the left (right) quantum well. 
The $L$ has an energy level of conduction electrons with up-spin, 
while 
the $R$ has 
$N_\uparrow$ energy levels of conduction electrons with up-spin 
and $N_\downarrow$, with down-spin. 
The shaded area in the electrodes represents 
the region occupied by electrons. 
The bias voltage (the gate voltage) is applied to the electrodes 
(the $L$ and the $R$). 
%Further, 
In addition, the equivalent circuit is shown in Fig. \ref{circuit}. 
%A voltage is applied to the left electrode. 
}
\label{model}
\end{center}
\end{figure}

In Fig. \ref{model}, 
we show a system consisting of 
``electrode/$L$/spin quantum dot/$R$/electrode" junctions, 
where 
the electrons flow from the left electrode to the right one 
under the bias voltage between the electrodes 
and furthermore 
the gate voltage is applied to the $L$ and the $R$ [see Sec. \ref{Appendix}]. 
Here, 
the dot region behaves as a tunnel barrier, 
and tunnel barriers are set 
between the left electrode and the $L$ 
and between the $R$ and the right electrode. 
%each ``/" represents a tunnel barrier. 
%The $L$ and $R$ therefore have the energy levels of the quantum well (QW). 
The $L$ and the $R$ form the QW; that is, 
%Moreover, the above barriers and ones of ``$L$/spin quantum dot/$R$" make the quantum wells (QWs) of the $L$ and $R$. 
%the tunnel barriers are set between the quantum wells (QWs) 
%and the electrodes and between the QWs and the dot. 
%In Fig. \ref{sw_dev1}, 
%These figures 
the $L$ has an energy level of conduction electrons with up-spin 
and becomes a spin filter to inject only the up-spin electrons, 
while the $R$ has 
$N_\uparrow$ energy levels of conduction electrons with up-spin 
and $N_\downarrow$, with down-spin. 
Those levels in the $R$ are introduced to accept electrons 
exhibiting the elastic and the inelastic transports 
[see Sec. \ref{assump}(iv)]. 
%In the case of $N_\uparrow$=0, the $R$ becomes the spin filter to accept only the down-spin electrons. 
%Both electrodes are composed of the nonmagnetic metal. 
%Note also that 
%Further, the barriers of ``electrode/$L$" and ``$R$/electrode" play an important role in bringing about the Coulomb blockade effect~\cite{CBE} [see Sec. \ref{assump}(i)]. 
%Both electrodes are nonmagnetic or 
The left electrode (right electrode) is 
a non-magnet or ferromagnet, 
in which the direction of the spin polarization is the same as 
that of the $L$ ($R$). 
%and 
The shaded area in the electrodes represents 
the region occupied by electrons. 
%%%The energy of the highest occupied orbitals   The energy of the most highly occupied state  the Fermi energy is the energy of the highest occupied state
%The energy of the highest occupied state is the Fermi energy, 
%It should be noted that 
%and some of electrons lying 
%between the Fermi level of the left electrode and that of the right one 
%contribute to the transport. 
In the dot, 
the quantum spin \mbox{\boldmath $S$}=$(S_x,S_y,S_z)$ 
with $S \ge 1$ is localized and 
%is localized in the dot and 
%and this spin 
is weakly coupled to the $L$ and the $R$. 
The localized spin has a uniaxial anisotropy energy 
showing a bistable potential, 
%$-|D|S_z^2$ with $-|D|$ being an anisotropy constant.~\cite{Mn12_1} 
$-|D|S_z^2$. 
The energy levels of the localized spin can be characterized by $S_z$. 
It is also assumed that 
the magnetic easy axis of the localized spin 
is collinear to 
that of the electron spins at the $L$ and the $R$. 
%the easy axis of electron spins at the $L$ and the $R$. 
In addition, 
%Furthermore, 
the magnetic couplings 
between the localized spin and the QWs are negligibly small. 
%, e.g.,in a magnetic .eld).

%When a voltage, $E_{\rm vol}$, is applied to the right electrode, 
%By applying a voltage, $E_{\rm vol}$, to the left electrode, the difference of height of highest occupied energy levels (HOELs) between the left and right electrodes appears as shown in Fig. \ref{model}. 
%and 
%we consider a situation in which 
%It is noted that 
%While the elctron runs through the dot, an exchange interaction acts between the electrons and the localized spin. 

%A voltage applied to the $L$, $E_{\rm vol}$, is here set to be $E_{\rm vol}$ = $|D| \left[ S^2 - (S-1)^2 \right]$=$|D|(2S-1)$, which is the maximum energy that the electron can give to the localized spin. 

\subsection{Assumptions}
\label{assump}

\begin{figure}[ht]
\begin{center}
\includegraphics[width=0.5\linewidth]{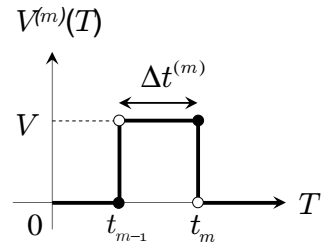}
\caption{
Interaction for the $m$th injected electron in the $m$th process, 
$V^{(m)}(T)$. 
%The potential profile for the $m$th injected electron in the $m$th process. 
Each process is independent. 
}
\label{poten}
\end{center}
\end{figure}

On condition that 
the initial state of the localized spin is $S_z$=$-S$, 
up-spin electrons are sequentially injected 
from the left electrode into the dot.~\cite{Eto} 
To describe the sequentially injected electrons, 
we have the following assumptions:~\cite{Eto} 
\begin{enumerate}
\item[(i)] 
%By applying a voltage to the left electrode, 
%A voltage is applied to the left electrode so that 
%This system exhibits the Coulomb blockade effect, 
%This system is in the Coulomb blockade regime, 
This system exhibits 
single electron tunneling (SET),~\cite{SET} 
in which the up-spin electrons are injected 
from the left electrode into the dot one by one 
%due to the Coulomb blockade effect 
under specific bias and gate voltages [see Sec. \ref{Appendix}]. 
%in which 
%; that is, more than one electron simultaneously cannot run through the dot because of the Coulomb repulsion among them.~\cite{CBE} 
%(Coulomb blockade effect). 
%in the presence of a voltage bias and in the Coulomb blockade regime 
%%%%As soon as an electron finishes going through the dot, the subsequent electron is injected into the dot and the previous electron moves more deeply into the right electrode. 
%this electron moves into the right electrode and 
%\hspace*{0.5cm}
%In other words, 
Concretely speaking, 
when the probability density of the electron becomes 0 at the $L$ 
and 1 at the $R$, 
the electron moves into the right electrode 
and the subsequent electron is injected into the $L$. 
%the sequential electron comes into the $L$. 
%and simultaneously the previous electron move into the right electrode. 
%\hspace*{0.2cm}
%To bring about the transport, 
%effect, 
%For such a transport, 
%We apply the bias and gate voltage to bring about this effect as shown in \ref{Appendix}. 
%$E_{\rm vol}$, is set to be $E_{\rm vol}$=$E_L + e^2/(4 \pi \epsilon r)$, 
%where 
%This $E_{\rm vol}$ corresponds to the Coulomb energy between them. 
%%%Under this $E_{\rm vol}$, as soon as the probability of the electron becomes 1 at the $R$, the subsequent electron comes into the $L$ and simultaneously the previous electron is pushed out of the $R$.
% and moves more deeply into the right electrode. 
%the electron at the $R$ and the sequential one of the left electrode. 
%Then, as soon as an electron finishes going through the dot, the subsequent electron is injected into the dot and the previous electron moves more deeply into the right electrode. 

\hspace*{0.2cm}
%As a notation, 
The situation in which the $m$th electron runs in the dot 
is named as the $m$th process. 
%When the transmission time of the $m$th electron 
When the time period of the $m$th process 
is represented by $\Delta t^{(m)}$, 
the finish time of the $m$th process, $t_m$, 
is defined by 
$t_m = \sum_{m'=1}^{m} \Delta t^{(m')}$. 
Here, $t_0$=0 is the initial time.

%set. 
\item[(ii)] 
When the transport of the electron 
is caused by interaction $V$ between the $L$ and the $R$,
$V$ acts while each electron runs in the $L$-dot-$R$ system. 
On the other hand, 
$V$ is switched off for the electron being outside of this system. 
%except for the motion in this system. 
%for each electron before the injection or after 
%each electron goes into this system or after it goes out of this system. 
The interaction 
for the $m$th injected electron at time $T$, $V^{(m)}(T)$, is written by, 
\begin{eqnarray}
\label{vvv}
{V}^{(m)}(T) =\left\{
\begin{array}{ll}
{V}, &\hspace{.5cm} {\rm for}~t_{m-1} < T \le t_m,
\\
0, &\hspace{.5cm} {\rm for}~T \le t_{m-1}, ~t_m < T, 
\end{array}
\right.
\end{eqnarray}
[see Fig. \ref{poten}]. 
As found from this $V^{(m)}(T)$, each process is independent. 
%Also, as described later, the spin flip transport of electrons gives rise to the reversal of the localized spin. 

%\subsection{Interaction $V$}
\hspace*{0.2cm}
The interaction $V$ is given 
by a transmission term 
including the exchange interaction 
between the electron and 
the localized spin.~\cite{Appelbaum1, Appelbaum2, Anderson, Kim} 
%We utilize the s-d exchange interaction as $V$. 
This term is obtained 
within the second-order perturbation theory 
based on the weak couplings between the dot and the QWs. 
The $L$, $R$, and dot parts are described by an 
unperturbed Hamiltonian, 
in which the on-site Coulomb energy is considered in the dot. 
The couplings between the dot and the QWs 
are perturbation terms. 
The resultant $V$ is now expanded to a more general system, 
namely, the anisotropic exchange interaction.~\cite{Shiba} 
The expression is written by,
\begin{eqnarray}
\label{int}
{V}&=&
V_0 \displaystyle{\sum_{n\uparrow=1}^{N_\uparrow}} \left( 
{c}_{\mbox{\tiny $R_{n\uparrow}$}}^\dag {c}_{\mbox{\tiny $L_\uparrow$}} +
{c}_{\mbox{\tiny $L_\uparrow$}}^\dag {c}_{\mbox{\tiny $R_{n\uparrow}$}}
\right) \nonumber \\
&&+J_\perp \displaystyle{\sum_{n\downarrow=1}^{N_\downarrow}} \left( 
{c}_{\mbox{\tiny $R_{n\downarrow}$}}^\dag {c}_{\mbox{\tiny $L_\uparrow$}} {S}_+  +
{c}_{\mbox{\tiny $L_\uparrow$}}^\dag {c}_{\mbox{\tiny $R_{n\downarrow}$}} {S}_-
\right) \nonumber \\
&&+J_z \displaystyle{\sum_{n\uparrow=1}^{N_\uparrow}} \left( 
{c}_{\mbox{\tiny $R_{n\uparrow}$}}^\dag {c}_{\mbox{\tiny $L_\uparrow$}}  +
{c}_{\mbox{\tiny $L_\uparrow$}}^\dag {c}_{\mbox{\tiny $R_{n\uparrow}$}}
\right) {S}_z. 
\end{eqnarray}
%with ${S}_\pm = {S}_x \pm i {S}_y$. 
Here, $c_{j}$ ($c_{j}^{\dag} $) is 
the annihilation (creation) operator of an electron 
%with $\sigma$ ($=\uparrow {\rm or} \downarrow$) spin 
of $j$=$L$ or $R_{n\sigma}$, 
where the suffix $L$ 
($R_{n\sigma}$) denotes the level of conduction electrons with up-spin 
of the $L$ 
(the $n$th level of conduction electrons with a $\sigma$ spin of the $R$). 
Furthermore, 
$V_0$ denotes the coefficient for direct tunneling, 
while $J_\perp$ ($J_z$) is that for tunneling with 
a transverse (longitudinal) exchange interaction 
between the electron and the localized spin. 
%Further, $J_\perp$ ($J_z$) is a transverse (longitudinal) exchange integral between the electron and the localized spin, and $V_0$ is a coefficient for the spin independent transport. 
From now on, we call $J_\perp$ ($J_z$) the exchange integral.

\item[(iii)]
The localized spin canted by the electron 
is not rapidly relaxed 
and interacts with the subsequently injected electron.

\item[(iv)]
%\label{as_iv}
%In addition, 
The system has the relation of $|E_L - E_{R_{n\sigma}}| \ll |V|$, 
where 
%When 
$E_L$ ($E_{R_{n\sigma}}$) denotes 
the energy level of conduction electrons with the up-spin of the $L$ 
(the $n$th energy level of conduction electrons 
with the $\sigma$ spin of the $R$). 
The energy levels contain the potential due to the gate voltage. 
%the present system is considered to have the relation of 
%$|E_L - E_{R_{n\sigma}}|$ is much smaller than the magnitude of $V$. 
%the exchange interaction. 
%The assumption is introduced to simplify the calculation 
%Here, 
%Furthermore, 
%Further, 
%Also, 
Here, 
the central region in $E_{R_{n\sigma}}$ is located 
in the vicinity of $E_L$. 
%and 
%the range of $E_{R_{n\sigma}}$ is from $-|D| (2S -1 )$ to $|D| (2S -1 )$, 
%the range is $-|D| (2S -1 ) \lesssim E_{R_{n\sigma}} \lesssim |D| (2S -1 )$, 
The band width in $E_{R_{n\sigma}}$ is about $2 |D| (2S -1 )$, 
with $|D| [S^2 - (S-1)^2]$=$|D| (2S -1 )$, 
where 
%the energy rage of $E_{R_{n\sigma}}$ 
%$-|D| (2S -1 ) \lesssim E_{R_{n\sigma}} \lesssim |D| (2S -1 )$ 
%that originates from 
%$-|D| (2S -1 )$ and 
$|D| (2S -1 )$ 
corresponds to 
%is related to 
the maximum energy that 
the conduction electron absorbs from the localized spin 
%and 
or gives to the spin 
%under 
for the case of the energy conservation. 
%the emission and the absorption of maximum energy for , respectively. 
%$-|D| [S^2 - (S-1)^2]$=$-|D| (2S -1 )$. 
%$|D| [S^2 - (S-1)^2]$=$|D| (2S -1 )$. 
%Therefore, 
Consequently, 
%The assumption of 
$|E_L - E_{R_{n\sigma}}| \ll |V|$
corresponds to $|D| \ll |V|$. 
%consequently means that $|D|$ is much smaller than the magnitude of $V$. 
%the exchange interaction. 
%$|D|$ is considerably small compared to the magnitude of the exchange interaction. 
%because 
%and 
Note that the assumption greatly simplifies the calculation 
[see Sec. \ref{sec_wave_fun}]. 

\hspace*{0.2cm}
As mentioned in Sec. \ref{sec_wave_fun} [also Ref. 24], 
%Furthermore, 
the above assumption leads to the relation of 
$\Delta t^{(m)} \ll \hbar/|E_L - E_{R_{n\sigma}}|$, 
%with $\hbar$ being the Planck constant $h$ divided by $2\pi$, 
where 
%$\Delta E$ corresponds roughly to 
%represents 
%the magnitude of difference 
%the magnitude of the largest difference 
%between energy levels of this system, and  
$\hbar$ is the Planck constant $h$ divided by $2\pi$. 
% [see Eq. (\ref{H_0})]. 
%the magnitude of energy change due to the transmission and $\hbar$ is the Planck constant $h$ divided by $2\pi$. 
%[see \ref{sec_wave_fun} and Ref. 23]. 
% and \cite{comment}]. 
%As a strategy, 
On the basis of the relation, 
we investigate the motion of the electron within the $L$-dot-$R$ system 
%and 
%although many electrons are sequentially injected into the dot. 
%We 
using the wave function obtained 
from the time-dependent Schr$\ddot{\rm o}$dinger equation. 
%on the basis of the relation of $\Delta t^{(m)} < 1/\Delta \omega$, 
%where $\Delta \omega$ represents the energy change due to the transmission. 
%At present, 
%Roughly speaking, 
%The present system is assumed to have 
%In this system, 
%$\Delta \omega$ is considered to be 
%$\Delta \omega \sim 10^{11}$ [1/s] and 
%$\Delta t^{(m)}$ is evaluated to be 
%$\Delta t^{(m)} \sim 10^{-13}$ [s]. 
%for this system. 
%for the $L$-dot-$R$ system. 
In contrast, 
the Fermi's golden rule 
which is a method to investigate the transport property, 
is applicable 
for the case of $\Delta t^{(m)} \gg \hbar/|E_L - E_{R_{n\sigma}}|$.

\hspace*{0.2cm}
% is reversed 
%owing to through s. 
%the localized spin is reversed 
%owing to 
%through the spin flip transport of the electrons. 
%Namely, 
%The electrons give and take energies with the localized spin, 
%and also 
% delta E \times delta t = hbar, where delta E = diference between the actual energy and the average of energy. 
We also note that 
in a short time period such as $\Delta t^{(m)}$, 
%the energy of the system is not always conserved 
the total energy 
%of the system 
becomes uncertain 
according to the uncertainty relation between time and energy, 
although the energy can be certainly conserved after enough time has passed. 
%Thus, 
The electrons then exhibit the elastic and the inelastic transports. 
%As for the latter, the electrons absorb (give) energies from (to) the localized spin system. 
% or they have emission of energies to the spin. 
%the electrons have absorption of energies from the localized spin or they have emission of energies to the spin. 
Such electrons are accepted by some energy levels in the $R$. 
% contribute to acceptances of . 
%Some energy levels in the $R$ contribute to acceptances of such electrons. 

\end{enumerate}

\section{Theoretical Formulation}
\label{theo_des}

\subsection{Wave function}
\label{sec_wave_fun}

In order to obtain the wave function of the $m$th process, 
$|\Phi^{(m)} (t) \rangle $ 
with $t \equiv T - t_{m-1}$ and $0 < t \le \Delta t^{(m)}$, 
we solve the time-dependent Schr$\ddot{\rm o}$dinger equation, 
% as follows: 
\begin{eqnarray}
\label{sch1}
i\hbar \frac{\partial }{\partial t} |\Phi^{(m)} (t) \rangle 
= {\cal H}^{(m)} (t)  |\Phi^{(m)} (t)  \rangle, 
%&&i\hbar \frac{\partial }{\partial t} |\phi^{(m)} (t)  \rangle 
%= {\cal H}_0 |\phi \rangle. 
\end{eqnarray}
with 
\begin{eqnarray}
{\cal H}^{(m)} (t)  = {\cal H}_0 + V^{(m)} (t), 
\end{eqnarray}
where ${\cal H}_0$ is the Hamiltonian for the electronic states 
of $L$, $R$, and the localized spin state. 
%Further, $\hbar$ is the Planck constant $h$ divided by $2\pi$. 
The wave function $|\Phi^{(m)} (t)  \rangle $ 
for $0 < t \le \Delta t^{(m)}$ is expressed as, 
\begin{eqnarray}
\label{wave_func}
|\Phi^{(m)} (t)  \rangle = \displaystyle{\sum_{j}\sum_{S_z=-S}^S} 
a_{j,S_z}^{(m)} (t) e^{-iE_{j,S_z} t/\hbar} |j, S_z \rangle,
\end{eqnarray}
with 
\begin{eqnarray}
\label{H_0}
{\cal H}_0 |j, S_z \rangle = E_{j,S_z} |j, S_z \rangle,
\end{eqnarray} 
for $j$=$L$ or $R_{n\sigma}$, and 
$E_{L,S_z}$=$E_L -|D|S_z^2$, 
$E_{R_{n\sigma},S_z}$=$E_{R_{n\sigma}} -|D|S_z^2$. 
%where $E_L$ ($E_{R_{n\sigma}}$) is the level of conduction electrons with up-spin of the $L$ (the $n$th level of conduction electrons with $\sigma$ spin of the $R$). 
The state $|L~(R_{n\sigma}),S_z \rangle$ means 
that the electron exists at the level of the $L$ 
(at the $n$th level of the $\sigma$ spin of the $R$) 
and the localized spin has $S_z$. 
The equation to determine the coefficient $a_{j,S_z}^{(m)} (t)$ 
for $0 < t \le \Delta t^{(m)}$ 
is given by, 
% as follows: 
%satisfies the following equation, 
\begin{eqnarray}
\label{a_k}
\dot{a}_{j,S_z}^{(m)} (t) 
&=& \displaystyle{\frac{1}{i \hbar} \sum_{k}\sum_{S_z'=-S}^S} 
\langle j,S_z | V | k, S_z' \rangle a_{k,S_z'}^{(m)} (t) \nonumber \\
&& \times e^{i (E_{j,S_z} -E_{k,S_z'})t/\hbar}. 
%V_{k,n} a_n (t) e^{i (\epsilon_k -\epsilon_n)t/\hbar}. 
\label{exact_a}
\end{eqnarray}
We obtain equations for 
$a_{L,S_z}^{(m)}(t)$, $a_{R_{n\uparrow},S_z}^{(m)}(t)$, 
and $a_{R_{n\downarrow},S_z+1}^{(m)}(t)$ 
using 
$\langle R_{n\downarrow}, S_z+1 | V | L, S_z \rangle 
= \langle L, S_z | V | R_{n\downarrow}, S_z+1 \rangle 
= J_\perp \sqrt{(S-S_z)(S+S_z+1)} \equiv F_{S_z}$ and 
$\langle R_{n\uparrow}, S_z | V | L, S_z \rangle = 
\langle L, S_z | V | R_{n\uparrow}, S_z \rangle = 
V_0 + J_z S_z \equiv G_{S_z}$. 
%We focus on an equation of them as follows: 
Their equations are summarized as follows: 
\begin{eqnarray}
\label{exact1}
&&\ddot{a}_{L,S_z}^{(m)} (t) =
- \frac{F_{S_z}^2}{\hbar^2} N_\downarrow a_{L,S_z}^{(m)} (t) 
- \frac{G_{S_z}^2}{\hbar^2} N_\uparrow a_{L,S_z}^{(m)} (t) \nonumber \\
&&\hspace*{1.6cm}+ f_{S_z}^{(m)}(t) + g_{S_z}^{(m)}(t), \\
\label{exact2}
&&\dot{a}_{R_{n\downarrow},S_z+1}^{(m)}(t) 
= -i\frac{F_{S_z}}{\hbar} a_{L,S_z}^{(m)} (t) 
e^{-i \Delta_{R_{n\downarrow},S_z} t/\hbar}, \\
\label{exact3}
&&\dot{a}_{R_{n\uparrow},S_z}^{(m)}(t) 
= -i\frac{G_{S_z}}{\hbar} a_{L,S_z}^{(m)} (t) 
e^{-i \Delta_{R_{n\uparrow},S_z} t/\hbar},
\end{eqnarray}
with 
\begin{eqnarray}
&&f_{S_z}^{(m)}(t)= \frac{F_{S_z}}{\hbar^2}
\sum_{n\downarrow=1}^{N_\downarrow} 
\Delta_{R_{n\downarrow},S_z} 
 a_{R_{n\downarrow},S_z+1}^{(m)}(t) 
e^{i \Delta_{R_{n\downarrow},S_z} t/\hbar}, \nonumber \\ 
&&g_{S_z}^{(m)}(t)= \frac{G_{S_z}}{\hbar^2}
\sum_{n\uparrow=1}^{N_\uparrow}
\Delta_{R_{n\uparrow},S_z} 
a_{R_{n\uparrow},S_z}^{(m)}(t)
e^{i \Delta_{R_{n\uparrow},S_z} t/\hbar}, \nonumber \\
\end{eqnarray}
$\Delta_{R_{n\downarrow},S_z} = E_{L,S_z}-E_{R_{n\downarrow},Sz+1}$, 
and $\Delta_{R_{n\uparrow},S_z} = E_{L,S_z}-E_{R_{n\uparrow},Sz}$, 
where 
the sum 
$\sum_{n\downarrow}$ ($\sum_{n\uparrow}$) 
%the sum of $f(t)$ [$g(t)$] 
excludes $n\hspace{-0.1cm}\downarrow$ 
($n\hspace{-0.1cm}\uparrow$) giving 
$\Delta_{R_{n\downarrow},S_z}$=0 
($\Delta_{R_{n\uparrow},S_z}$=0). 
%The dash in the sum means that the value ko = 0 is excluded

%We assume here that 
%We here consider a case in which 
%We now focus on a case in which the present system has the condition, 
%On the basis of \ref{as_iv}, 
%we focus on a system with a condition, 
According to 
$|E_L - E_{R_{n\sigma}}| \ll |J_\perp|$, $|J_z|$, and $|V_0|$ of Sec. \ref{assump}(iv), 
the present system has the relation of 
$|\Delta_{R_{n\downarrow},S_z}/ F_{S_z}| \ll 1$ and 
$|\Delta_{R_{n\uparrow},S_z}/ G_{S_z}| \ll 1$.~\cite{comment} 
%are taken into account, 
%meaning that the energy level of $L$ is close to the energy levels of $R$.~\cite{comment} 
%{\it i.e.} $E_{\rm vol} + E_L \sim E_{R_{n\sigma}}$, 
%the energy level of $L$, $E_{\rm vol} + E_L$, is close to 
%the energy levels of $R$,  $E_{R_{n\sigma}}$, 
%the energy levels of $R$ are close to the energy level of $L$ 
%and further $|D|$ is much smaller than $|V_0|$, $|J_\perp|$, and $|J_z|$. 
%Then, 
The time average of $|f_{S_z}^{(m)}(t)|$ and 
$|g_{S_z}^{(m)}(t)|$ then becomes negligibly small 
compared to that of the magnitude of the first and second terms 
on the right-hand side of Eq. (\ref{exact1}) 
because 
the sum in the complex plane is done in addition to this relation 
and $|a_{R_{n\downarrow},S_z+1}^{(m)}(t)|$ 
and $|a_{R_{n\uparrow},S_z}^{(m)}(t)|$ are 1 at maximum. 
We hence obtain $a_{L,S_z}^{(m)}(t)$ from an approximate expression 
by omitting $f_{S_z}^{(m)}(t)$ and $g_{S_z}^{(m)}(t)$ from Eq. (\ref{exact1}).

The assumption for $a_{L,S_z}^{(m)}(t)$ of Sec. \ref{assump}(i) 
is concretely defined as follows: 
under the voltages described in Secs. \ref{assump}(i) and \ref{Appendix}, 
as soon as the local probability density 
at the $L$ of the $m$th injected electron becomes zero, 
the $m$th electron goes out of the $R$, 
and the potential for the $m$th electron is then switched off: 
$V^{(m)}(T)=0$ for $T > t_m$. 
Next, the subsequent $(m+1)$th electron moves 
%is injected 
%from the deep inside of the left electrode to the $L$. 
from the left electrode to the $L$. 
%the dot. 
%Simultaneously, 
%Because of the Coulomb repulsion from the $(m+1)$th electron, 
%The $m$th electron is pushed out of the $R$ under the influence of the Coulomb repulsion from the $(m+1)$th electron, 
The initial condition of the $m$th process is therefore given by, 
\begin{eqnarray}
\label{condition1}
a_{L,S_z}^{(m)}(0) &=& 
\left[ \sum_{n\uparrow=1}^{N_\uparrow} 
\left| 
\langle R_{n\uparrow}, S_z | \Phi^{(m-1)}(\Delta t^{(m-1)}) \rangle \right|^2  
\right. \nonumber \\
&& \left. 
+ \sum_{n\downarrow=1}^{N_\downarrow} 
\left| 
\langle R_{n\downarrow}, S_z | \Phi^{(m-1)}(\Delta t^{(m-1)}) \rangle 
\right|^2 \right]^{1/2} \nonumber \\
%&=& \left| \langle R_{n,\sigma}, S_z | 
%\Phi^{(m-1)}(\Delta t^{(m-1)}) \rangle \right| \nonumber \\
&\equiv& b_{S_z}^{(m)}, \\
\label{condition2}
\dot{a}_{L,S_z}^{(m)}(0)&=&0. 
\end{eqnarray}
Here, $b_{S_z}^{(m)}$ is 
the initial amplitude of the $m$th process with $0 \le b_{S_z}^{(m)} \le 1$. 
It is represented by using the probability amplitude 
specified by 
the final state of the localized spin of the $(m-1)$th process 
according to Sec. \ref{assump}(iii). 
In particular, 
$b_{-S}^{(1)}$=1 is apparent due to $S_z$=$-S$ at $T$=0. 
%$b_{-S}^{(1)}$=1 is set reflecting $S_z$=$-S$ at $T$=0. 
Furthermore, $\dot{a}_{L,S_z}^{(m)}(0)$=0 originates from 
$V^{(m)}(T)$=0 for $T \le t_{m-1}$ of Eq. (\ref{vvv}). 
%in Sec. \ref{assump}(ii). 

As a result, the coefficients are obtained as follows: 
\begin{eqnarray}
\label{aaa}
&&a_{L,S_z}^{(m)}(t)=b_{S_z}^{(m)} \cos (\Omega_{S_z} t),  \\
\label{aaa1}
&&a_{R_{n\downarrow},S_z+1}^{(m)}(t) 
=  -\frac{iF_{S_z}b_{S_z}^{(m)}}{\hbar \Omega_{S_z}} \sin (\Omega_{S_z} t), \\
\label{aaa2}
&&a_{R_{n\uparrow},S_z}^{(m)}(t) 
=  -\frac{iG_{S_z}b_{S_z}^{(m)}}{\hbar \Omega_{S_z}} \sin (\Omega_{S_z} t), 
\end{eqnarray}
with 
\begin{eqnarray}
\label{omega_Sz}
\Omega_{S_z} = \frac{\sqrt{N_\downarrow F_{S_z}^2 + N_\uparrow G_{S_z}^2}}
{\hbar}, 
\end{eqnarray}
for $0 < t \le \Delta t^{(m)}$. 
Here, 
$a_{R_{n\downarrow},S_z+1}^{(m)}(t)$ 
and $a_{R_{n\uparrow},S_z}^{(m)}(t)$ are obtained by using 
$a_{L,S_z}^{(m)}(t)$ and Eqs. (\ref{exact2}) and (\ref{exact3}), 
where 
$\Delta_{R_{n\downarrow},S_z}t/\hbar $ and 
$\Delta_{R_{n\uparrow},S_z}t/\hbar $ 
are taken as 0 
because $t$ is considered to be, at most, several times larger than 
$1/\Omega_{S_z}$ in this study~\cite{comment1} and further 
$|\Delta_{R_{n\downarrow},S_z}/ F_{S_z}| \ll 1$ 
and $|\Delta_{R_{n\uparrow},S_z}/ G_{S_z}| \ll 1$ are introduced 
as stated earlier.

%Actually, 
In fact, the coefficients of Eqs. (\ref{aaa}), (\ref{aaa1}), and (\ref{aaa2}) 
correspond to exact expressions 
%which are obtained 
for a system of $D$=0 and 
$\Delta_{R_{n\downarrow},S_z}$=$\Delta_{R_{n\uparrow},S_z}$=0 
($E_L=E_{R_{n\uparrow}}=E_{R_{n\downarrow}}$) 
obtained under the condition that 
%when 
the initial state of the localized spin is set to be $S_z$=$-S$. 
%From a viewpoint of investigations of the qualitative behaviors, 
%we believe that 
%In investigating the qualitative behaviors, 
%In investigation of 
%When the qualitative behaviors are investigated, 
%the use of 
%in order to 
%For the purpose of invesitageting the qualitative behaviors. 
%We believe, however, that 
%However, 
The use of the coefficients, however, is considered to be 
%believed to be 
valid 
for the study of the qualitative properties 
%is considered to be 
%considered to be 
%valid for 
%are qualitatively same as those of 
%we investigate the qualitative behaviors 
%of the spin reversal for 
of the present system with 
%$|D| \ll |V_0|$, $|J_\perp|$, $|J_z|$, 
%and 
$|E_L - E_{R_{n\sigma}}| \ll |J_\perp|$, $|J_z|$, and $|V_0|$. 
%in order to qualitatively invesitagete the properties. 

%in which the energy levels of $R$ are close to the energy level of $L$ and $|D|$ is much smaller than $|V_0|$, $|J_\perp|$, and $|J_z|$. 
%meaning that the energy levels of $R$ are close to the energy level of $L$ and $|D| \ll |V_0|$, $|J_\perp|$, and $|J_z|$. 
%$\Delta_{R_{n\downarrow},S_z}/ F_{S_z} \ll 1$ 
%and $\Delta_{R_{n\uparrow},S_z}/ G_{S_z} \ll 1$. 

We substitute Eqs. (\ref{aaa}), (\ref{aaa1}), and (\ref{aaa2}) 
into Eq. (\ref{wave_func}). 
The wave function $|\Phi^{(m)}(t) \rangle$ 
for $0 < t \le \Delta t^{(m)}$ is finally written as, 
\begin{eqnarray}
\label{wf}
&&|\Phi^{(m)}(t) \rangle = 
\displaystyle{\sum_{S_z=-S}^{S}}
A_{L,S_z}^{(m)} (t) e^{-iE_{L,S_z} t/\hbar} |L,S_z \rangle \nonumber \\
&&+ 
\displaystyle{\sum_{n\uparrow=1}^{N_\uparrow}} 
\displaystyle{\sum_{S_z=-S}^{S}}
A_{R_{n\uparrow},S_z}^{(m)} (t) 
e^{-iE_{R_{n\uparrow},S_z} t/\hbar}
 |R_{n\uparrow},S_z \rangle  \nonumber \\
&& + 
\displaystyle{\sum_{n\downarrow=1}^{N_\downarrow}}
\displaystyle{\sum_{S_z=-S}^{S-1}}
A_{R_{n\downarrow},S_z+1}^{(m)} (t) 
e^{-iE_{R_{n\downarrow},S_z+1} t/\hbar} |R_{n\downarrow},S_z+1 \rangle, \nonumber \\
\end{eqnarray}
with 
%\begin{eqnarray}
$\displaystyle{ A_{L,S_z}^{(m)}(t) 
= \frac{a_{L,S_z}^{(m)}(t) }{a^{(m)}(t)}}$, 
$\displaystyle{A_{R_{n\uparrow},S_z}^{(m)}(t) 
= \frac{a_{R_{n\uparrow},S_z}^{(m)}(t) }{a^{(m)}(t)}}$, 
$\displaystyle{A_{R_{n\downarrow},S_z+1}^{(m)}(t) 
= \frac{a_{R_{n\downarrow},S_z+1}^{(m)}(t) }{a^{(m)}(t)}}$, 
%\end{eqnarray}
where $a^{(m)}(t)$ is the normalization factor, 
\begin{eqnarray}
a^{(m)}(t) &=& \left[ 
\sum_{S_z=-S}^{S}
\left| a_{L,S_z}^{(m)}(t) \right|^2
+ 
%\sum_{n\uparrow=1}^{N_\uparrow} 
N_\uparrow 
\sum_{S_z=-S}^{S}
\left| a_{R_{n\uparrow},S_z}^{(m)} (t) \right|^2 
\right. 
\nonumber \\
&&\left. 
+ 
%\sum_{n\downarrow=1}^{N_\downarrow} 
N_\downarrow 
\sum_{S_z=-S}^{S-1}
\left| a_{R_{n\downarrow},S_z+1}^{(m)} (t) \right|^2 
\right]^{1/2}. 
\end{eqnarray}
The local probability densities at the $L$ and the $R$ are 
given by, 
\begin{eqnarray}
\label{LLL}
\langle \Phi^{(m)}(t) | {\cal L} |\Phi^{(m)}(t) \rangle  &=& 
%\frac{1}{2S+1}
\sum_{S_z=-S}^{S}
\left| A_{L,S_z}^{(m)}(t) \right|^2, \\
\label{RRR}
\langle \Phi^{(m)}(t) | {\cal R} |\Phi^{(m)}(t) \rangle  &=& 
%\frac{1}{N_\uparrow (2S+1) + 2N_\downarrow S} \times  
%\displaystyle{\sum_{n\uparrow=1}^{N_\uparrow}} 
N_\uparrow
\sum_{S_z=-S}^{S} 
\left|A_{R_{n\uparrow},S_z}^{(m)}(t)\right|^2  \nonumber \\
%&&+\displaystyle{\sum_{n\downarrow=1}^{N_\downarrow}} 
&& \hspace*{-1cm}+ N_\downarrow 
\sum_{S_z=-S}^{S-1}
\left|A_{R_{n\downarrow},S_z+1}^{(m)}(t)\right|^2, 
\end{eqnarray}
respectively, with 
$ {\cal L} \equiv 
%(2S+1)^{-1/2}
\sum_{S_z=-S}^{S} 
| L,S_z \rangle \langle L, S_z |$ and 
$ {\cal R} \equiv 
%\left[ N_\uparrow (2S+1) + 2N_\downarrow S \right]^{-1/2}
\sum_{n\uparrow=1}^{N_\uparrow}
\sum_{S_z=-S}^{S}
|R_{n\uparrow},S_z \rangle \langle R_{n\uparrow},S_z | + 
\sum_{n\downarrow=1}^{N_\downarrow}
\sum_{S_z=-S}^{S-1}
|R_{n\downarrow},S_z+1 \rangle \langle R_{n\downarrow},S_z+1 |$.

\subsection{Number of electrons in the $R$ and the right electrode}

%Based on 
On the basis of Eq. (\ref{RRR}), we obtain 
the number of electrons in the $R$ and the right electrode 
for the $m$th process, 
$n_r^{(m)} (t)$. 
The expression is written by,
\begin{eqnarray}
\label{probability}
n_r^{(m)} (t) &=& (m-1) + 
%\displaystyle{\sum_{n\uparrow=1}^{N_\uparrow}} 
N_\uparrow 
\sum_{S_z=-S}^{S} 
\left|A_{R_{n\uparrow},S_z}^{(m)}(t)\right|^2 \nonumber \\
&&+
N_\downarrow 
%\displaystyle{\sum_{n\downarrow=1}^{N_\downarrow}} 
\sum_{S_z=-S}^{S-1}
\left|A_{R_{n\downarrow},S_z+1}^{(m)}(t)\right|^2, 
\end{eqnarray}
for $0 < t \le \Delta t^{(m)}$. 
The first term in the right-hand side represents the number of electrons 
which have moved to the right electrode. 
The second and third terms are 
$\langle \Phi^{(m)}(t) | {\cal R} |\Phi^{(m)}(t) \rangle$ 
of Eq. (\ref{RRR}). 
%the local probability density 
%at the $R$ of the $m$th injected electron, 
%which is given by Eq. (\ref{RRR}). 
%Here, 
Furthermore, $n_r^{(m)} (t) - (m-1)
~\left[ = \langle \Phi^{(m)}(t) | {\cal R} |\Phi^{(m)}(t) \rangle \right]$ 
is regarded as 
the expectation value of the position of the $m$th injected electron 
when the positions of the $L$ and the $R$ are set to be 0 and 1, respectively.

%\subsection{Velocity}
\subsection{Electrical current}

The electrical current of the $m$th electron, $I^{(m)} (t)$, is defined by 
%$I(t) = \sum_{m} I^{(m)}$=$-e \sum_{m} v^{(m)} (t)$, with 
$I^{(m)}(t)$=$-e n_e S_d v^{(m)}(t)$, where 
$v^{(m)}(t)$ is the velocity of the $m$th electron, 
$e$ ($>$0) is the electric charge, 
$n_e$ is the number density of electrons, and 
$S_d$ is the cross-sectional area of the dot. 
The velocity, 
%of the $m$th electron, 
%process, 
$v^{(m)}(t)$, is obtained from 
the time differential of 
the expectation value of the position of the $m$th injected electron. 
When the positions of the $L$ and the $R$ are 
defined as 0 and $l$, respectively, 
%the electrical current of the $m$ electron, 
$I^{(m)}(t)$ is written as follows: 
%that is, 
%the local probability density at the $L$. 
%The expression 
%which is written by, 
%When the positive direction is defined to be from the $L$ to the $R$, 
%the expression is written as follows: 
\begin{eqnarray}
\label{current}
I^{(m)}(t) 
&=& -e n_e S_d l \frac{d}{dt} 
\langle \Phi^{(m)}(t) | {\cal R} |\Phi^{(m)}(t) \rangle \nonumber \\
%%\left[ n_r^{(m)} (t) - (m-1) \right] \nonumber \\
%\hspace*{-1cm}&=& \frac{d}{dt} \left[ N_\uparrow \sum_{S_z=-S}^{S} \left|A_{R_{n\uparrow},S_z}^{(m)}(t)\right|^2  + N_\downarrow \sum_{S_z=-S}^{S-1} \left|A_{R_{n\downarrow},S_z+1}^{(m)}(t)\right|^2 \right] \nonumber \\
%&=& -e \frac{d}{dt} l \left[ 1-\langle \Phi^{(m)}(t) | {\cal L} |\Phi^{(m)}(t) \rangle \right] \nonumber \\
%&=& 
%-\frac{d}{dt} \sum_{S_z = -S}^S \left| A_{L,S_z}^{(m)} (t) \right|^2  
%\nonumber \\
%=-\frac{d}{dt} \sum_{S_z = -S}^S(b_{S_z}^{(m)})^2 \cos^2 (\omega_{S_z} t) 
&=& \frac{-e n_e S_d l}{\left[ a^{(m)}(t) \right]^2}
\sum_{S_z = -S}^S \left[ b_{S_z}^{(m)} \right]^2 \Omega_{S_z} 
\sin (2 \Omega_{S_z} t), \nonumber \\
\end{eqnarray}
for $0 < t \le \Delta t^{(m)}$. 
%Here, $\ell$ represents the distance between the $L$ and the $R$. 
%A minus sign is put in front of this time differential 
%because the time differential represents outflow from the $L$. 
%Here, $\left| (R |\Phi^{(m)}(t) \rangle \right|^2 + \left| ( L |\Phi^{(m)}(t) \rangle \right|^2 = 1$ is taken into account. 

\subsection{Expectation value of the localized spin}

%we obtain 
The expectation value of the localized spin of the $m$th process, 
$\langle S_z^{(m)} (t) \rangle$, 
is obtained by using Eq. (\ref{wf}). 
%$\langle S_z^{(m)} (t) \rangle = 
%\langle \Phi^{(m)}(t) |S_z |\Phi^{(m)}(t) \rangle$, 
The expression is given by, 
%as follows: 
\begin{eqnarray}
\label{av_sz}
\langle S_z^{(m)} (t) \rangle &=& 
\langle \Phi^{(m)}(t) | {S}_z |\Phi^{(m)}(t) \rangle \nonumber \\
&& \hspace*{-2cm}= \sum_{S_z=-S}^{S}
S_z \left|A_{L,S_z}^{(m)}(t)\right|^2  
%\nonumber\\
%&&\hspace*{-0.5cm}
+
N_\uparrow 
%\displaystyle{\sum_{n\uparrow=1}^{N_\uparrow}} 
\sum_{S_z=-S}^{S} 
S_z 
\left|A_{R_{n\uparrow},S_z}^{(m)}(t)\right|^2 \nonumber \\
&&\hspace*{-1.7cm}+
N_\downarrow 
%\displaystyle{\sum_{n\downarrow=1}^{N_\downarrow}} 
\sum_{S_z=-S}^{S-1}
(S_z + 1) 
\left|A_{R_{n\downarrow},S_z+1}^{(m)}(t)\right|^2, 
\end{eqnarray}
for $0 < t \le \Delta t^{(m)}$.

%\section{Application to the case of spin-polarized $R$ 
\section{Applications}
\subsection{Spin-polarized $R$ with only down-spin levels} 
\label{appl1}

The spin reversal occurs 
when the $R$ has energy levels 
of conduction electrons with only down-spin. 
In this system, $V$ is described by 
%corresponds to 
only the second term in Eq. (\ref{int}) 
because of $N_\uparrow$=0. 
Furthermore, $G_{S_z}$, $a_{R_{n\uparrow}, S_z}^{(m)}(t)$, 
and $A_{R_{n\uparrow}, S_z}^{(m)}(t)$ 
are set to be 0.

%\subsection{Wave function}
%\label{sec_wave}

\begin{figure}[ht]
\begin{center}
\includegraphics[width=0.58\linewidth]{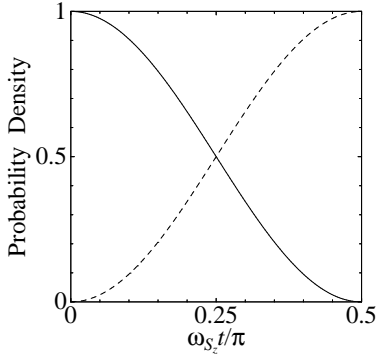}
\caption{
Time $t$ dependence of the partial probability density of the $m$th process. 
Here, 
$| \langle L,S_z |\Phi^{(m)}(t) \rangle |^2$ 
($\sum_{n\downarrow=1}^{N_\downarrow}
| \langle R_{n\downarrow},S_z+1 
|\Phi^{(m)}(t) \rangle |^2$) is shown 
by a solid (dashed) curve. 
}
\label{pro_den}
\end{center}
\end{figure}

%Using eqs. (\ref{wf}), (\ref{aa1}), and (\ref{aa2}), we obtain $|\Phi^{(m)}(t) \rangle$ for $0 < t \le \Delta t^{(m)}$. 
We obtain 
$|\Phi^{(m)}(t) \rangle$ for $0 < t \le \Delta t^{(m)}$ 
using 
Eqs. (\ref{aaa}), (\ref{aaa1}), (\ref{wf}), 
and $b_{S_z}^{(m)}$= $\delta_{m,S_z+S+1}$ 
which is Eq. (\ref{condition1}) for this system. 
%which is obtained in this system. 
The expression is written by, 
\begin{eqnarray}
\label{wf_spR}
&&|\Phi^{(m)}(t) \rangle = 
\cos (\omega_{S_z} t) e^{-iE_{L,S_z} t/\hbar} |L,S_z \rangle  \nonumber \\
&&-\sum_{n\downarrow=1}^{N_\downarrow} 
%\frac{i}{\sqrt{N_\downarrow}} 
\frac{i J_\perp}{\sqrt{N_\downarrow} |J_\perp|} \sin (\omega_{S_z} t)
e^{-iE_{R_{n\downarrow},S_z+1} t/\hbar} |R_{n\downarrow},S_z+1 \rangle, \nonumber \\
\end{eqnarray}
with 
\begin{eqnarray}
\label{om_rev}
\omega_{S_z} =
\frac{\left| J_\perp \right| \sqrt{N_\downarrow (S-S_z)(S+S_z+1)}}{\hbar}. 
\end{eqnarray}
and $m$=$S_z + S + 1$ for $S_z$=$-S$ - $S-1$. 
%In the following, we consider the partial probability density, 
From Eq. (\ref{wf_spR}), 
the partial probability density is obtained as follows: 
\begin{eqnarray}
\label{A_L}
&&\left| \langle L,S_z |\Phi^{(m)}(t) \rangle \right|^2 = 
%\left| A_{L,S_z}^{(m)}(t) \right|^2 = 
\cos^2 (\omega_{S_z}t), \\
\label{A_Rn}
&&\sum_{n\downarrow=1}^{N_\downarrow}
\left| \langle R_{n\downarrow},S_z+1 
|\Phi^{(m)}(t) \rangle \right|^2 
%\sum_{n\downarrow=1}^{N_\downarrow}
%\left| A_{R_{n\downarrow},S_z+1}^{(m)}(t) \right|^2 
=\sin^2 (\omega_{S_z}t), 
\end{eqnarray}
with $m$=$S_z + S + 1$ for $S_z$=$-S$ - $S-1$. 
%with $m=S_z + S + 1$.  
For $m=S_z+S+1$, 
$\left| \langle L,S_z |\Phi^{(m)}(t) \rangle \right|^2$ 
decreases from 1 to 0 
and 
$\sum_{n\downarrow=1}^{N_\downarrow}
\left| \langle R_{n\downarrow},S_z+1 
|\Phi^{(m)}(t) \rangle \right|^2$ increases from 0 to 1 
with increasing $t$ [see Fig. \ref{pro_den}]. 
%Namely, 
This behavior means that the electron transfers from the $L$ to the $R$ 
with the spin flip from up to down 
and the localized spin changes from $S_z$ to $S_z + 1$. 
%[for details, see \ref{sec_n_r} and \ref{sec_Sz}]. 
%as will be mentioned in \ref{sec_n_r} and \ref{sec_Sz}. 
%It is found that 
Since the $m$th process is finished at $t=\pi/(2 \omega_{S_z})$, 
$\Delta t^{(m)}$ is given by, 
%corresponds to $\Delta t^{(m)}$ with $m = S_z + S + 1$~($m$ =1 - $2S$). 
\begin{eqnarray}
\label{delta_sz}
\Delta t^{(m)} =
\frac{\pi}{2 \omega_{S_z}} =
\frac{h}{4\left| J_\perp \right| \sqrt{N_\downarrow (S-S_z)(S+S_z+1)}}, 
\end{eqnarray}
with $m = S_z + S + 1$ for $S_z$=$-S$ - $S-1$. 
%~($m$ =1 - $2S$). 
%Note that 
Here, 
%the relation of 
$m = S_z + S + 1$ for $S_z$=$-S$ - $S-1$ represents 
%the process number 
%$m$ is related to one value of $S_z$; that is, 
$m$=1, 2, 3, 4, $\cdot \cdot \cdot$, 2$S$. 
%It is mentioned that $m$ of the $m$th process has one to one correspondence to $S_z$. 
The spin reversal is thereby achieved after the ($2S$)th process, 
where $2S$ also corresponds to the total number of injected electrons. 
%as will be mentioned in \ref{sec_n_r}. 

%\subsection{Reversal time of localized spin}
%\label{sec_rev_t}

\begin{figure}[ht]
\begin{center}
\includegraphics[width=0.65\linewidth]{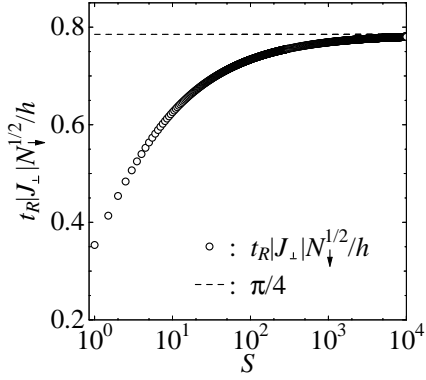}
\caption{
Spin quantum number $S$ dependence of 
the reversal time of the localized spin, $t_{R}$. 
%\times |J_\perp| \sqrt{N_\downarrow}/h$. 
Here, 
$t_{R} |J_\perp| \sqrt{N_\downarrow}/h$ 
takes $\sqrt{2}/4$ at $S$=1, while 
it comes close to $\pi/4$ in the classical spin limit of $S \to \infty$. 
}
\label{t_r}
\end{center}
\end{figure}

Using Eq. (\ref{delta_sz}), 
we can obtain the reversal time of the localized spin, $t_{R}$, 
which is the total time 
to transform $S_z$ from $-S$ to $S$. 
When $\Delta t^{(m)}$ with $m = S_z + S + 1$ of Eq. (\ref{delta_sz}) 
is renamed as $\Delta t_{\mbox{\footnotesize $S_z$}}$, 
%$\Delta t_{\mbox{\footnotesize $S_z$}}$ becomes 
$t_{R}$ is simply written by,
\begin{eqnarray}
\label{rev_time}
t_{R} &=& \displaystyle{\sum_{S_z = -S}^{S-1} 
\Delta t_{\mbox{\footnotesize $S_z$}}} \nonumber \\
&=& \displaystyle{\sum_{S_z = -S}^{S-1} 
%h/\left[4|J_\perp| \sqrt{N_\downarrow (S-S_z)(S+S_z+1)} \right] }. 
\frac{h}{4 \left| J_\perp \right| \sqrt{N_\downarrow (S-S_z)(S+S_z+1)}}}. 
\end{eqnarray}
Figure \ref{t_r} shows the $S$ dependence of 
$t_{R}  |J_\perp| \sqrt{N_\downarrow}/h$. 
At $S$=1, $t_R$ takes $h\sqrt{2}/(4|J_\perp| \sqrt{N_\downarrow})$. 
On the condition that $|J_\perp| \sqrt{N_\downarrow}/h$ is a constant, 
%It is found that 
%it is shown that 
%the reversal time 
$t_{R}$ monotonically increases with increasing $S$. 
This behavior is explained by considering that 
the spin reversal is caused 
by injecting $2S$ up-spin electrons into the dot. 
%Further, 
%Moreover, 
In the classical spin limit of $S \to \infty$, 
$t_{R}$ 
%finally 
comes close to $h\pi/(4|J_\perp| \sqrt{N_\downarrow})$ 
which is obtained by replacing $\sum_{S_z = -S}^{S-1}$ 
%in Eq. (\ref{t_r}) 
with $\lim_{S \to \infty} \int_{-S}^{S-1}dS_z$. 
%At present, 
%Note also that 
We thus find that 
the $t_R$ of systems with $S \ge 1$ has 
the following relation, 
\begin{eqnarray}
\frac{h\sqrt{2}}{4|J_\perp| \sqrt{N_\downarrow}} 
\le t_R < \frac{h\pi}{4|J_\perp| \sqrt{N_\downarrow}}. 
\end{eqnarray}
Furthermore, since 
this $t_{R}$ is derived under Sec. \ref{assump}(iii), 
%, in which the localized spin is not rapidly relaxed. 
we emphasize that the spin reversal is realized 
when the relaxation time of the localized spin, $\tau$, 
satisfies the relation of 
$\tau > t_{R}
= \sum_{S_z = -S}^{S-1} h/\left[ 4|J_\perp| 
\sqrt{N_\downarrow (S-S_z)(S+S_z+1)} \right]$.

As an application, we consider a system with 
$S$=5, $N_\downarrow$=50, and $|J_\perp|$=0.001 eV. 
%For the applied system in \ref{sec_wave}, 
The reversal time, $t_{R}$, is evaluated to be 3.3 $\times$ 10$^{-13}$ s 
by using Eq. (\ref{rev_time}). 
%In this system, 
The relation of $ \tau > 3.3 \times 10^{-13}$ s 
is essential for an experimental observation of this spin reversal. 
%In this case, the system with $\tau$ satisfying $ \tau > 3.3 \times 10^{-13}$ [s] is essential for an experimental observation of the reversal. 

%%This value of $t_R$ is found also from Fig. \ref{s5}. 

\begin{figure}[ht]
\begin{center}
\includegraphics[width=0.65\linewidth]{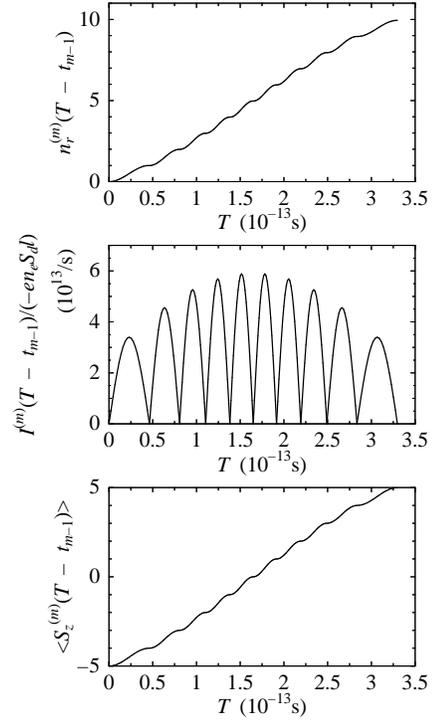}
\caption{
Time $T$ dependence of 
the number of electrons in the $R$ and the right electrode, 
$n_r^{(m)} (T- t_{m-1})$; 
electrical current, $I^{(m)}(T- t_{m-1})$; 
and 
expectation value of the localized spin, 
$\langle S_z^{(m)}(T - t_{m-1}) \rangle$; 
$m$=1 - 10 
for the system of $S$=5, $N_\downarrow$=50, and $|J_\perp|$=0.001 eV. 
Upper panel: $n_r^{(m)} (T- t_{m-1})$. 
Middle panel: $I^{(m)}(T- t_{m-1})/(-e n_e S_d l)$. 
Lower panel: $\langle S_z^{(m)}(T - t_{m-1}) \rangle$. 
}
\label{s5}
\end{center}
\end{figure}

%As an application, we consider a system with 
%%in the case of 
%$S$=5, $N_\downarrow$=50, and $|J_\perp|$=0.001 eV. 
The upper panel of Fig. \ref{s5} shows the $T$ dependence of 
$n_r^{(m)} (T -t_{m-1})$ with $m$=1 - 10. 
%for the applied system in \ref{sec_rev_t}. 
The number of electrons in the $R$ and the right electrode, 
$n_r^{(m)}(T -t_{m-1})$, increases from 0 to 10 with increasing $T$ 
and becomes 10 at $T=t_R=3.3 \times 10^{-13}$ s. 
%where 
The reversal is eventually achieved by 
injecting 10 (=$2S$) electrons into the dot. 
Furthermore, 
%it is characteristic that the slope 
it is noted that the slope 
becomes zero at $t_m$ 
with $m$=1 - 9.

The middle panel of Fig. \ref{s5} shows 
the $T$ dependence of 
$I^{(m)}(T -t_{m-1})/(-e n_e S_d l)$. 
% for the applied system in \ref{sec_rev_t}. 
%$1/\Delta t_{\mbox{\footnotesize $S_z$}}$ 
%for $S$=5, in the case of $N$ =50. 
%is plotted for the elapsed time 
%$t_{\mbox{\footnotesize $S_z$}}~(=\sum_{S_z=-S}^{S_z-1})$, 
%where $t_{\mbox{\footnotesize $S_z$}}$ means 
%the total time to transform the localized spin from $-S$ to $S_z$. 
%The electrical current 
The quantity $I^{(m)}(T -t_{m-1})/(-e n_e S_d l)$ exhibits 
a single peak for each process. 
The peak takes the largest value for the 5th and 6th processes, 
%$m$=5 and 6, 
where 
%the initial states in their processes 
the respective initial states have $S_z$ = $-$1 and 0. 
%which corresponds to $S_z$ = $-1$, 0, respectively. 
This behavior reflects the fact that 
$\omega_{S_z}$ 
% \propto \sqrt{N_\downarrow (S-S_z)(S+S_z+1)}$ 
%of Eq. (\ref{om_rev}) 
achieves its maximum at $S_z$ = $-$1 and 0 for $S$=5 
[see also Eq. (\ref{current})]. 
%At each $t_{\mbox{\footnotesize $S_z$}}$, $1/\Delta t_{\mbox{\footnotesize $S_z$}}$ becomes large with increasing $S$. In the case of large $S$, the magnitude of interaction $V$ becomes large, and the conduction electron flow faster. 

The lower panel of Fig. \ref{s5} 
shows the $T$ dependence of $\langle S_z^{(m)} (T -t_{m-1}) \rangle$. 
%for the applied system in \ref{sec_rev_t}. 
%with $m$=1 - 10, in the case of $S$=5, $N_\downarrow$=50, and $|J_\perp|$=0.001 [eV]. 
The expectation value of the localized spin 
$\langle S_z^{(m)}(T -t_{m-1}) \rangle $ 
increases from $-$5 to 5 with increasing $T$ 
following the same pattern as that of $n_r^{(m)} (T -t_{m-1})$. 
It finally becomes 5 at $T=t_R=3.3 \times 10^{-13}$ s; that is, 
the spin reversal is completed at this $t_R$. 
%Furthermore, 
%it is characteristic that the slope 
%we note that the slope 
%becomes zero at $t_m$ 
%with $m$=1 - 9. 
%Furthermore, $\displaystyle{\frac{d }{d t} \langle S_z^{(m)}(t) \rangle } $=0 is obviously found at each $\displaystyle{t= \frac{\pi}{2 \omega_{S_z}}}$.   

\begin{figure}[ht]
\begin{center}
\includegraphics[width=0.65\linewidth]{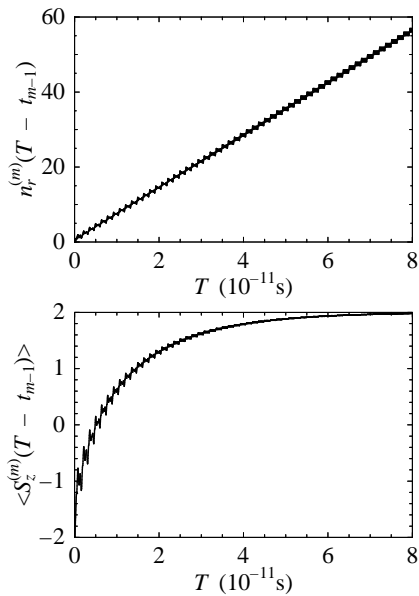}
\caption{
Time $T$ dependence of 
$n_r^{(m)} (T- t_{m-1})$ and 
$\langle S_z^{(m)}(T - t_{m-1}) \rangle$ with $m$=1 - 57 
for the system of $S$=2 and $N_\uparrow$=$N_\downarrow$=20. 
The parameter set of $V$ is 
$V_0$=0.001 eV, $J_\perp/V_0$=12/31, and $J_z/V_0=2\sqrt{11}/31$. 
Upper panel: $n_r^{(m)} (T- t_{m-1})$. 
Lower panel: $\langle S_z^{(m)}(T - t_{m-1}) \rangle$. 
}
\label{reverse1}
\end{center}
\end{figure}

\begin{figure}[ht]
\begin{center}
\includegraphics[width=0.65\linewidth]{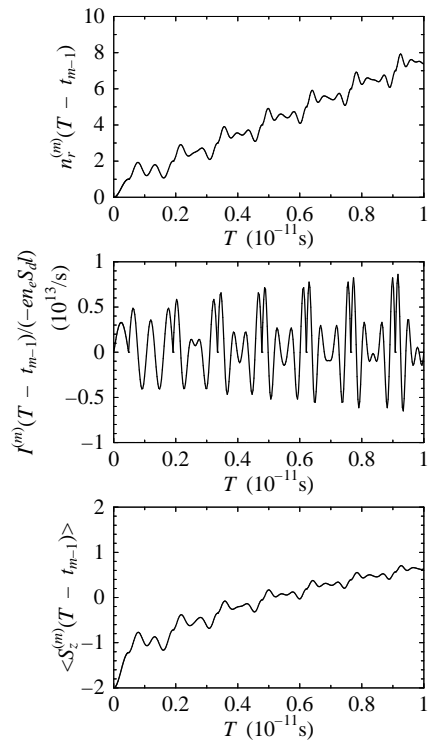}
\caption{
%that is, 
Time $T$ dependence of 
$n_r^{(m)} (T- t_{m-1})$, 
$I^{(m)}(T- t_{m-1})/(-e n_e S_d l)$, 
and $\langle S_z^{(m)}(T - t_{m-1}) \rangle$ with $m$=1 - 8 
for the system of $S$=2 and $N_\uparrow$=$N_\downarrow$=20. 
The parameter set of $V$ 
is the same as that of Fig. \ref{reverse1}. 
Upper panel: $n_r^{(m)} (T- t_{m-1})$. 
Middle panel: $I^{(m)}(T- t_{m-1})/(-e n_e S_d l)$. 
Lower panel: $\langle S_z^{(m)}(T - t_{m-1}) \rangle$. 
Here, 
$n_r^{(m)} (T- t_{m-1})$ and $\langle S_z^{(m)}(T - t_{m-1}) \rangle$ 
are the respective ones on the larger scale of Fig. \ref{reverse1}. 
}
\label{reverse2}
\end{center}
\end{figure}

\subsection{Non-magnetic $R$} 
%\section{Application to the case of nonmagnetic $R$}
\label{appl2}

In the case of non-magnetic $R$ with $N_\uparrow$=$N_\downarrow$, 
%where the $R$ 
%which has energy levels of conduction electrons with equal number of up and down-spins ($N_\uparrow$=$N_\downarrow$), 
this system exhibits spin reversal or non-reversal 
depending on the exchange integral. 
%%We show the both phenomena. 
% with the other $S$. 
%with the other $S$. 

%In this system, $V$ is precisely given by Eq. (\ref{int}) because of $N_\uparrow \ne$ 0 and $N_\downarrow \ne$ 0. Since $V$ includes the term with $S_z$, $b_{S_z}^{(m)}$ takes finite values for various $S_z$ when $m \ge 2$. In other words, the initial state of the $m$th process with $m \ge 2$ consists of states with various $S_z$. 

%Moreover, $n_r^{(m)} (t)$, $v^{(m)}(t)$, $\langle S_z^{(m)}(t) \rangle$ are identical to eqs. (\ref{probability}), (\ref{velocity}), (\ref{av_sz}), respectively. 

%\subsection{Local probability density at the $L$}

%We first investigate a parameter set 
%which can induce the spin reversal. 
%for $S$=2 and $N_\uparrow$=$N_\downarrow$=20. 
%The spin reversal may occur, 
%In the case of the spin reversal, 
Regarding the spin reversal, 
%when all injected electrons can transmit 
all injected electrons can transfer 
from the $L$ to the $R$; that is, 
the local probability density at the $L$ 
becomes zero at a certain time for each process [see Sec. \ref{assump}(i)]. 
We therefore pay attention to 
the local probability density 
at the $L$ of the $m$th process, 
\begin{eqnarray}
\label{local_L}
%\frac{1}{2S+1}
%\displaystyle{\sum_{S_z = -S}^{S}}
\langle \Phi^{(m)}(t) | {\cal L} |\Phi^{(m)}(t) \rangle 
%= \left| A_{L,S_z}^{(m)} (t) \right|^2 
\propto 
%\frac{1}{2S+1}
\displaystyle{\sum_{S_z = -S}^{S}}
\left[ b_{S_z}^{(m)} \right]^2 \cos^2 (\Omega_{S_z} t), 
\end{eqnarray}
which is obtained from Eqs. (\ref{aaa}) and (\ref{LLL}). 
%for $S_z$ = $-$2, $-$1, 0, 1, 2. 
%When Eq. (\ref{local_L}) 
%%$\cos^2 (\omega_{S_z} t)$ for each $S_z$ 
%becomes zero at a certain time for each process, 
%the injected electrons can transmit from the $L$ to the $R$ in all processes. 
%Equation (\ref{local_L}) takes zero 
Equation (\ref{local_L}) becomes zero at $t=\Delta t^{(m)}$, 
when $\Delta t^{(m)}$ satisfies the following relation, 
\begin{eqnarray}
\label{om_relate}
\Omega_{S_z} \Delta t^{(m)} = \left( n_{S_z}^{(m)} + \frac{1}{2} \right) \pi, 
\end{eqnarray}
%is satisfied 
for any $S_z$ giving finite $b_{S_z}$. 
Here, 
%$\omega_{S_z}$ is Eq. (\ref{omega_Sz}) and 
$n_{S_z}^{(m)}$ $(\ge 0)$ is the counting number.

In the first process with $b_{-S}^{(1)}$=1, 
Eq. (\ref{local_L}) becomes 
$\cos^2 (\Omega_{-S} t)$, 
and it certainly takes zero 
% satisfies 
at $t= \Delta t^{(1)} = \pi/(2 \Omega_{-S})$. 
%$ \Delta t^{(1)} = \pi/(2\omega_{S_z})$. 
%$n_{-S}$=0. 
%In the $m$th process with $m \ge 5$, 
%On the other hand, 
%In the $m$th process with $m \ge 2$,  
%$b_{S_z}^{(m)}$ is finite for any $S_z$. 
%=-2, -1, 0, 1, 2$. 
%Therefore, it is necessary to satisfy 
In the $m$th process with $m \ge 2$, 
$b_{S_z}^{(m)}$ is finite for any $S_z$. 
Whether Eq. (\ref{om_relate}) is satisfied for $m \ge 2$ 
%the spin reversal occurs 
or not depends on 
the parameter set of the exchange integral.

\begin{table}
\caption{
Initial amplitude of the $m$th process, 
$b_{S_z}^{(m)}$; 
counting number of the $m$th process, $n_{S_z}^{(m)}$; 
and time period of the $m$th process, $\Delta t^{(m)}$ 
for the system of $S$=2, $N_\uparrow$=$N_\downarrow$=20, 
%The parameter set, 
$V_0$=0.001 eV, $J_\perp/V_0$=12/31, and $J_z/V_0=2\sqrt{11}/31$. 
%, is used. 
}
  \label{tab1}
  \begin{tabular}{ccccccc} \\ \hline
$m$ &  $b_{-2}^{(m)}$   & $b_{-1}^{(m)}$ &  $b_{0}^{(m)}$ &  $b_{1}^{(m)}$ &  $b_{2}^{(m)}$ & $\Delta t^{(m)}$ \\ 
    &  $n_{-2}^{(m)}$   & $n_{-1}^{(m)}$ &  $n_{0}^{(m)}$ &  $n_{1}^{(m)}$ &  $n_{2}^{(m)}$ & \\ 
\hline
  1  & 1 & 0  & 0 & 0 & 0 & $\displaystyle{\frac{\pi}{2 \Omega_{-2}}}$   \\
     & 0 & 0  & 0 & 0 & 0 &     \\
  2  & 0.47 & 0.88 & 0 & 0 & 0  & $\displaystyle{\frac{3\pi}{2 \Omega_{-2}}}$     \\
     & 1 & 2  & 0 & 0 & 0      \\
  3  & 0.22 & 0.79 & 0.57 & 0 & 0 & $\displaystyle{\frac{3\pi}{2 \Omega_{-2}}}$      \\ 
     & 1 & 2  & 3 & 0 & 0      \\
  4  & 0.10  & 0.64 & 0.72 & 0.26 & 0 & $\displaystyle{\frac{3\pi}{2 \Omega_{-2}}}$      \\ 
     & 1 & 2  & 3 & 4 & 0      \\
  5  & 0.50 $\times 10^{-1}$ & 0.50 & 0.76 & 0.41 & 0.73$\times 10^{-1}$  & $\displaystyle{\frac{3\pi}{2 \Omega_{-2}}}$      \\ 
     & 1 & 2  & 3 & 4 & 5      \\
$\cdot$ & $\cdot$ & $\cdot$ & $\cdot$ & $\cdot$ & $\cdot$ & $\cdot$ \\[-.25cm] 
$\cdot$ & $\cdot$ & $\cdot$ & $\cdot$ & $\cdot$ & $\cdot$ & $\cdot$ \\[-.25cm] 
$\cdot$ & $\cdot$ & $\cdot$ & $\cdot$ & $\cdot$ & $\cdot$ & $\cdot$ \\ 
  30  & 0.38 $\times 10^{-9}$ & 0.65 $\times 10^{-3}$ & 0.58$\times 10^{-1}$ & 0.44 & 0.90 & $\displaystyle{\frac{3\pi}{2 \Omega_{-2}}}$      \\ 
     & 1 & 2  & 3 & 4 & 5      \\
$\cdot$ & $\cdot$ & $\cdot$ & $\cdot$ & $\cdot$ & $\cdot$ & $\cdot$ \\[-.25cm] 
$\cdot$ & $\cdot$ & $\cdot$ & $\cdot$ & $\cdot$ & $\cdot$ & $\cdot$ \\[-.25cm] 
$\cdot$ & $\cdot$ & $\cdot$ & $\cdot$ & $\cdot$ & $\cdot$ & $\cdot$ \\ 
  57  & 0.67 $\times 10^{-18}$& 0.50$\times 10^{-6}$ & 0.26$\times 10^{-2}$ & 0.14 & 0.99 & $\displaystyle{\frac{3\pi}{2 \Omega_{-2}}}$      \\ 
     & 1 & 2  & 3 & 4 & 5      \\
\hline
  \end{tabular}
\end{table}

%\subsection{Spin reversal}

%In particular, the $m$th process with $m \ge S$ has 
%$b_{S_z}^{(m)} \ne 0$ for $S_z=-S, -S+1, \cdot \cdot \cdot, S$. 
%Equation (\ref{local_L}) therefore takes zero 
%when $\omega_{S_z} \Delta t^{(m)} = (n_{S_z} + 1/2 ) \pi$ 
%is satisfied for $S_z=-S, -S+1, \cdot \cdot \cdot, S$, 
%where 
%$\Delta t^{(m)}$ is independent of $S_z$ and 
%$n_{S_z}$ is the counting number. 

We first investigate the parameter set satisfying Eq. (\ref{om_relate}) 
in order to find the spin reversal. 
%which can induce the spin reversal. 
%for $S$=2 and $N_\uparrow$=$N_\downarrow$=20. 
For a system with $S$=2 and $N_\uparrow$=$N_\downarrow$=20, 
%for $S$=2 and $N_\uparrow$=$N_\downarrow$=20. 
%The first process has $t= \Delta t^{(1)}=\pi/(2 \omega_{-2})$ . 
%The first process with $b_{-S}^{(1)}$=1 has $n_{-S}$=0. 
%On the other hand, the $m$th process with $m \ge 5$ has 
%$b_{S_z}^{(m)} \ne 0$ with $S_z=-2, -1, 0, 1, 2$. 
%Therefore, it is necessary to satisfy 
% at the same $\Delta t^{(m)}$, 
%where $\Delta t^{(m)}$ is independent of $S_z$ and 
%$n_{S_z}$ is the counting number. 
%Note here that the first process with $b_{-S}^{(1)}$=1 has only $n_{-S}$=0. 
%The finish time of the $m$th process, $t_m$, is then given by,
%\begin{eqnarray}
%t_m = t_1 + (m-1) \Delta t
%\end{eqnarray}
%with $t_1=\pi/(2 \omega_{-2})$ and 
%$\Delta t=(n_{S_z} + 1/2 ) \pi/\omega_{S_z}$. 
the parameter set 
%for the spin reversal 
is, for example, 
$V_0$=0.001 eV, $J_\perp/V_0$=12/31, $J_z/V_0=2\sqrt{11}/31$. 
%satisfying Eq. (\ref{om_relate}). 
%We $\omega_{S_z} \Delta t^{(m)} = (n_{S_z} + 1/2 ) \pi$ 
%is satisfied for $S_z=-S, -S+1, \cdot \cdot \cdot, S$, 
%Here, the first process with $b_{-2}^{(1)}$=1 has $n_{-2}$=0. 
In Table \ref{tab1}, 
we summarize $b_{S_z}^{(m)}$, $n_{S_z}^{(m)}$, and $\Delta t^{(m)}$ 
for each $m$. 
%It is noted that 
The first process with $b_{-2}^{(1)}$=1 has 
$\Delta t^{(1)} = \pi/(2 \Omega_{-2})= 4.78 \times 10^{-13}$ s, 
%%$n_{-2}$=0. 
and the $m$th process with $m \ge 2$ has 
$\Delta t^{(m)} = 3 \pi/(2 \Omega_{-2})=1.43 \times 10^{-12}$ s. 
%1.911194775270111E-012 - 4.777986938175278E-013 = 1.433396E-12

%In the upper panel of Fig. \ref{reverse1}, we show $n_r^{(m)}(t)$ with the parameter set and $m$=1 - 57. 
The upper panel of Fig. \ref{reverse1} 
and the upper panel of Fig. \ref{reverse2} 
show the $T$ dependence of $n_r^{(m)}(T -t_{m-1})$ 
with the above parameter set 
in the regions of $0 \le T \le 8 \times 10^{-11}$ s 
and $0 \le T \le 1 \times 10^{-11}$ s, respectively. 
Here, $T = 8 \times 10^{-11}$ s ($T = 1 \times 10^{-11}$ s) 
corresponds to the time in the 57th (8th) process. 
%in regions of $0 \le T \le 8 \times 10^{-11}$ [s] and $0 \le T \le 1 \times 10^{-11}$ [s], respectively. 
The number of electrons in the $R$ and the right electrode, 
$n_r^{(m)}(T -t_{m-1})$, 
slightly oscillates in each process 
and clearly increases with increasing $T$. 
%As found from 
In the middle panel of Fig. \ref{reverse2}, 
it is shown that 
$I^{(m)}(T -t_{m-1})/(-e n_e S_d l)$ oscillates 
between positive and negative values
%and takes positive and negative values. 

The lower panel of Fig. \ref{reverse1} and 
the lower panel of Fig. \ref{reverse2} 
show 
the $T$ dependence of $\langle S_z^{(m)}(T -t_{m-1}) \rangle$ 
in the respective regions of $T$. 
%with the parameter set and $m$=1 - 57. 
The expectation value of the localized spin, 
$\langle S_z^{(m)}(T -t_{m-1}) \rangle$, 
has a slight oscillation in each process as well. 
% and totally increases with increasing $T$. 
The spin reversal is almost realized 
in the vicinity of $T=8.0 \times 10^{-11}$ s, 
and it 
%This behavior 
is also confirmed 
from $b_{2}^{(57)} = 0.99$ in Table \ref{tab1}.

\begin{figure}[ht]
\begin{center}
\includegraphics[width=0.65\linewidth]{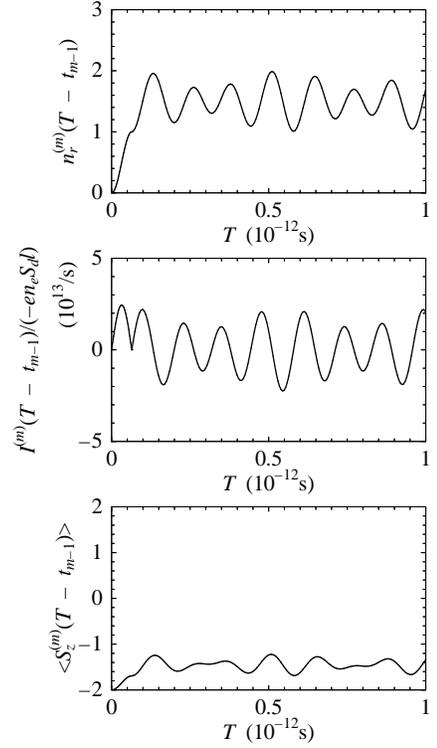}
\caption{
Time $T$ dependence of 
$n_r^{(m)} (T- t_{m-1})$, 
$I^{(m)}(T- t_{m-1})/(-e n_e S_d l)$, 
and $\langle S_z^{(m)}(T - t_{m-1}) \rangle$ 
with $m$=1, 2 
for the system of $S$=2 and $N_\uparrow$=$N_\downarrow$=20. 
The parameter set of $V$ 
is $V_0$=0.001 eV, $J_\perp/V_0$=1, and $J_z/V_0=2$. 
Upper panel: $n_r^{(m)} (T- t_{m-1})$. 
Middle panel: $I^{(m)}(T- t_{m-1})/(-e n_e S_d l)$. 
Lower panel: $\langle S_z^{(m)}(T - t_{m-1}) \rangle$. 
}
\label{trap}
\end{center}
\end{figure}

%\subsection{Spin nonreversal}

As the spin non-reversal case for $S$=2 and $N_\uparrow$=$N_\downarrow$=20, 
we choose, for example, the parameter set of 
$V_0$=0.001 eV, $J_\perp/V_0$=1, $J_z/V_0$=2, 
which does not have $t$ to satisfy Eq. (\ref{om_relate})
%$\omega_{S_z} t = (n_{S_z} + 1/2 ) \pi$ 
for $S_z=-2, -1, 0, 1, 2$. 
% [see Eq. (\ref{om_relate})]. 
% at the same $t$. 
%It is noted however that the first injected electron is transmittable. 
In the first process with $b_{-2}^{(1)}$=1, 
%however, 
%we obtain 
%Namely, we obtain 
Eq. (\ref{local_L}) becomes 
$
%\sum_{S_z = -2}^{2}
\left| \langle \Phi^{(1)}(t) | {\cal L} |\Phi^{(1)}(t) \rangle \right|^2 
\propto 
%\displaystyle{\sum_{S_z = -S}^{S}}
\cos^2 (\Omega_{-2} t)$ 
and it takes zero at 
$t= \pi/(2 \Omega_{-2})$=$6.4 \times 10^{-14}$ s. 
%$| A_{L,-2}^{(1)} ( \pi/(2 \omega_{-2}) ) |^2$=0 
%with $\pi/(2 \omega_{-2})$=$6.4 \times 10^{-14}$ [s]. 
In fact, the first injected electron is transferable 
as seen from the upper panel of Fig. \ref{trap}. 
In the second process, 
%$|A_{L,-1}^{(2)}(t)|^2
%$(2S+1)^{-1}\sum_{S_z = -S}^{S}\left| \langle L, S_z |\Phi^{(2)}(t) \rangle \right|^2$ 
Eq. (\ref{local_L}) always has finite values. 
%%The secondly injected electron is therefore trapped in the dot 
The second injected electron is trapped in the dot 
and it oscillates 
between the $L$ and the $R$ 
as shown in the upper and middle panels of Fig. \ref{trap}. 
The localized spin also oscillates between $S_z$=$-2$ and $S_z$=$-1$ 
[see the lower panel of Fig. \ref{trap}]. 
%It is also noted that the qualitatively same behaviors are found for systems with the other $S$. 

In addition, 
we report that 
qualitative behaviors identical to those noted above are found 
for some systems (some $S$'s). 
%In addition, we mention that 
%the qualitatively same behaviors as the above ones are found 
%confirmed 
%for some systems (some $S$'s). 

\section{Comments}
\label{comments}

%Based on the present study, 
%On the basis of the present study, 
%Based on the above results, 
%we give 
%a speculation about observations 
%and a proposal towards device applications. 
%the following comments. 
%towards device applications. 

%In addition, 
%\begin{enumerate}
%\item[(1)] 
%We frist 
%First, we give 
What follows is a comment about 
%consider 
an application of 
%have a speculation for 
the system with the non-magnetic $R$. 
According to the present study, 
the system exhibits a spin reversal or non-reversal 
%whether the spin reversal is realized 
%the spin reversal occurs 
%or not 
depending on the exchange integral. 
The exchange integral is originally described by 
the orbital energy level and the on-site Coulomb energy in the dot 
and transfer integrals 
between the dot and the QWs.~\cite{Schrieffer} 
Thereby, 
%when the orbital energy level is easily controlled by applying the gate voltage to the dot, 
by controlling the orbital energy level 
with the application of the gate voltage, 
this system may be 
switchable between the spin reversal and the spin non-reversal.

\begin{figure}[ht]
\begin{center}
\includegraphics[width=0.76\linewidth]{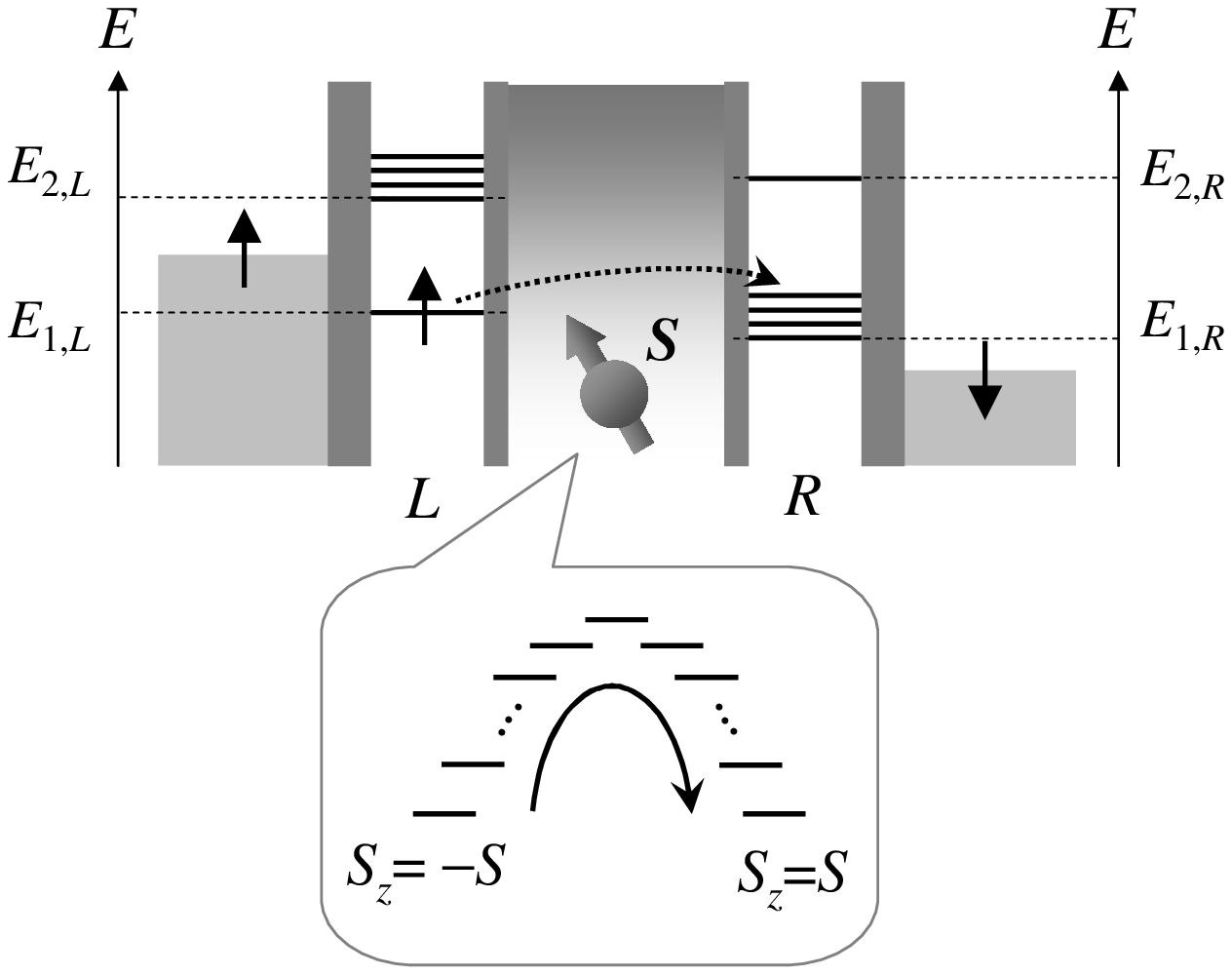}\\
\hspace*{-7.5cm}(a)\\
\vspace{0.2cm}
\includegraphics[width=0.76\linewidth]{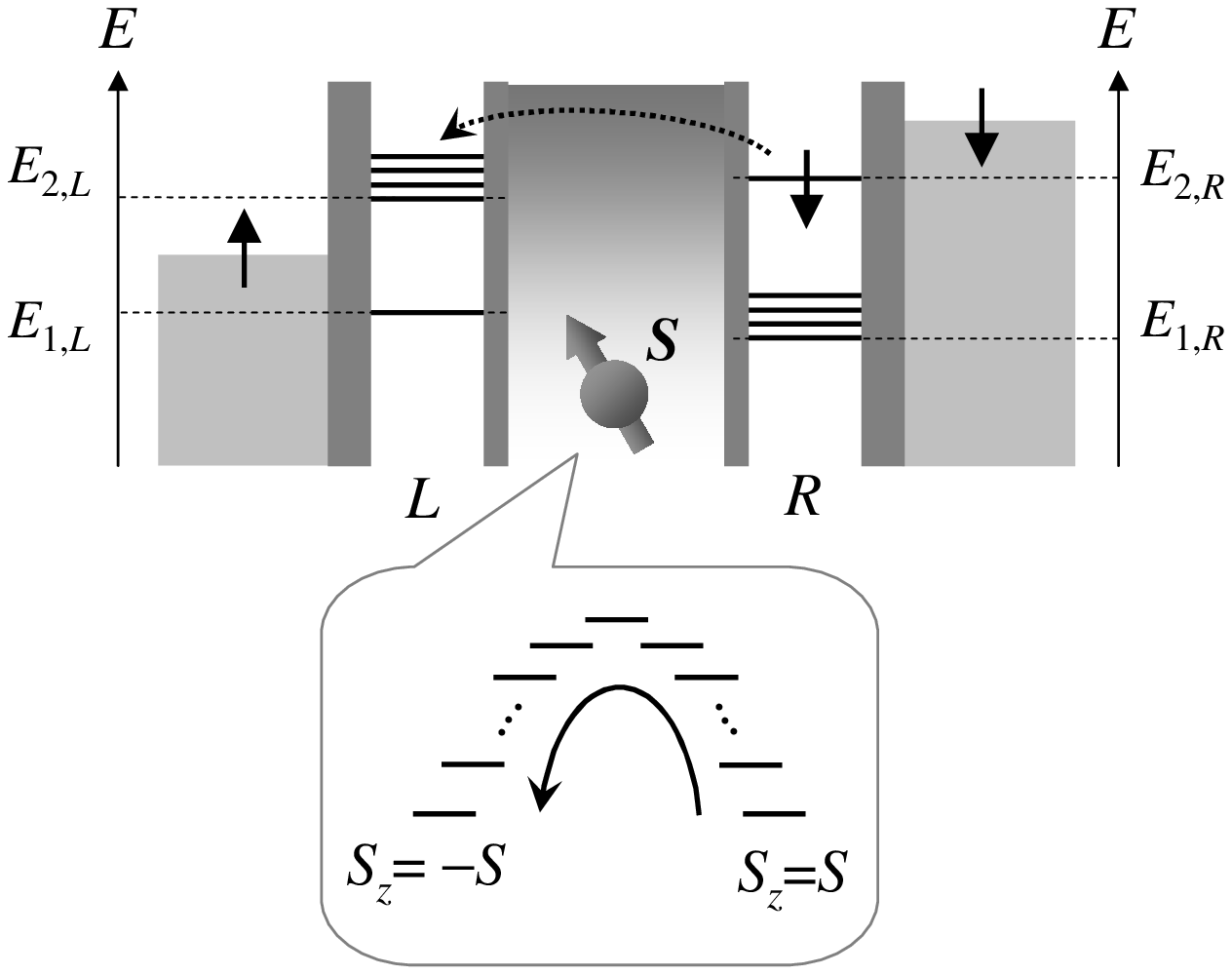}\\
\hspace*{-7.5cm}(b)
\caption{
A model with reversible switching 
% the reversible switching 
between $S_z$=$-S$ and $S$. 
%i.e. electrode/$L$/dot/$R$/electrode junctions. 
%In the vicinity of $E_{1,L}$, 
% and $E_{1,R}$, 
%the $L$ has an energy level of conduction electrons with up-spin, 
%while the $R$ has 
%$N_\downarrow$ energy levels of conduction electrons with down-spin. 
%and $N_\downarrow$ ones with down-spin. 
%In the vicinity of $E_{2,R}$, 
% and $E_{2,L}$, 
%the $R$ has an energy level of conduction electrons with down-spin, 
%while the $L$ has 
%$N_\uparrow'$ energy levels of conduction electrons with up-spin. 
%and $N_\downarrow'$ ones with down-spin. 
(a) The reversal from $S_z$=$-S$ to $S$ is induced 
by injecting the up-spin electrons from the $L$ into the dot. 
(b) The reversal from $S_z$=$S$ to $-S$ is achieved 
by injecting the down-spin electrons from the $R$ into the dot. 
}
\label{sw_dev1}
\end{center}
\end{figure}

%\item[(2)] 
%As a proposal
%Based on the present study, 
Secondly, we propose 
%we show 
%a system exhibiting 
a model with reversible switching 
between $S_z$=$-S$ and $S$ [see Fig. \ref{sw_dev1}]. 
%depending on 
%due to the direction of the spin injection. 
% and the spin direction. 
% towards device applications. 
%, based on the present study. 
%in Figs. \ref{sw_dev1} and \ref{sw_dev2} 
%[see Figs \ref{sw_dev1} and \ref{sw_dev2}]. 
%In Fig. \ref{sw_dev1}, 
%These figures 
%we show the model consisting of ``electrode/spin-polarized layer/dot/spin-polarized layer/electrode" junctions, where the tunnel barriers are set between the layers and the electrodes and between the layers and the dot. 
%in which 
%occupied regions of electrons. 
%energy regions 
%in which electrons are occupied. 
%Here, 
%%%%   The left and right layers are now named by the $L$ and the $R$, respectively. 
%and 
%At the $L$, 
The $L$ has energy levels of conduction electrons with up-spin, 
in which 
an energy level 
%is located 
is located at $E_{1,L}$ 
%and bottom of higher levels are done 
and the lowest of the $N_\uparrow'$ energy levels is found at $E_{2,L}$. 
%, where $E_{2,L}$ is higher than $E_{1,L}$. 
The $R$ has levels of conduction electrons with down-spin, 
in which 
%$N_\downarrow$ energy levels of conduction electrons with down-spin. 
%where . 
%At the $R$, on the other hand, 
a level is located at $E_{2,R}$ and 
the lowest of the $N_\downarrow$ levels lies at $E_{1,R}$. 
Here, tunneling probabilities between the level at $E_{1,L}$ 
of the $L$ and that at $E_{2,R}$ of the $R$ and 
between the levels in the vicinity of $E_{2,L}$ of the $L$ 
and those in the vicinity of $E_{1,R}$ of the $R$ 
are assumed to be negligibly small. 
%     a level (bottom of $N_\downarrow$ levels) is located at $E_{2,R}$ ($E_{1,R}$). 
%and bottom of $N_\downarrow$ levels is at $E_{1,R}$. 
%and lower levels are in the vicinity of 
%In particular, the electrons in the region of $E < E_{1,R}$ ($E < E_{2,L}$) of Fig. \ref{sw_dev1} (a) (Fig. \ref{sw_dev1} (b)) have no contribution to the transport. 
%\hspace*{0.5cm}
%For a lower energy region 
%    The shaded area in the electrodes represents the region occupied by electrons. 
%The Fermi level of the right electrode 
The energy of the highest occupied state (i.e. the Fermi level) 
of the right electrode 
is changed by applying the bias voltage, 
while that of the left electrode is fixed 
%at $E_{3,L}$ which is located 
between $E_{1,L}$ and $E_{2,L}$. 
%Note that 
It should be noted that some electrons lying 
%between the Fermi level of the left electrode 
between the energy of the highest occupied state of the left electrode 
and that of the right electrode contribute to the transport. 
%It is noted that 
%%the energy structures of the $L$ and the $R$ 
%the transport in the vicinity of $E_{1,L}$ 
%are considered to be those of the system of Sec. \ref{appl1}. 
%; that is, 
%where 
%in which 
%the $L$ has an energy level of conduction electrons with up-spin, 
%while the $R$ has 
%$N_\downarrow$ energy levels of conduction electrons with down-spin. 
%and $N_\downarrow$ ones with down-spin. 
%The situation of Fig. \ref{sw_dev1} (a) 
%corresponds just to that of Fig. \ref{model}. 
%shows 
%Namely, 
%a case in which the conduction electron in the vicinity of $E_{1,L}$ is injected from the $L$ into the dot. 
%In the vicinity of $E_{1,L}$ and $E_{1,R}$, 
%This case 
%energy structures of the $L$ and $R$ 
%corresponds just to the system of Fig. \ref{model}. 
%when 

We now consider the SET regime as described in Sec. \ref{Appendix}. 
When the bias voltage 
%(the voltage of the left electrode) 
is applied to the right electrode 
as shown in Fig. \ref{sw_dev1} (a), 
%(the voltage of the right electrode) 
%is now applied to the right electrode 
%so as to meet 
%the energy level of the $R$, 
%so that the Fermi level becomes $E_{3,R}$ 
%which satisfies 
%$E_{3,R}$=$E_{3,L} - [E_{1,L} + e^2/(4\pi \epsilon r)]$ 
%with $r_{LR} < r < r_0$ based on \ref{assump}(i) 
%as shown in  Fig. \ref{sw_dev1} (a). 
%and the energy level of the left electrode is set to be $E_2$, 
the up-spin electrons are injected from the $L$ into the dot 
%[see Fig. \ref{sw_dev1} (a)]. 
% and then 
%As a result, 
and the reversal from $S_z$=$-S$ to $S$ can be induced 
through the interaction of Eq. (\ref{int}) with ${N_\uparrow}$=0. 
%by injecting the up-spin electron from the $L$ into the $R$ 
%[see Fig. \ref{sw_dev1} (a)]. 
%When the voltage of the left electrode is applied as as to meet 
%the energy level of the $L$, 
%$E_1$,  
%and the energy level of the right electrode is set to be $E_1 - E_{\rm vol}$, 
%the reversal from $S_z$=$-S$ to $S$ is induced 
%by injecting the up-spin electron 
%from the $L$ into the $R$. 
%\hspace*{0.5cm}
%On the other hand, 
%For a higher energy region 
On the other hand, in the vicinity of $E_{2,R}$, 
% and $E_{2,L}$, 
%the number of the levels and the spin direction 
%are inversed opposite configuration to the case in the vicinity of $E_1$. 
%Namely, 
%%the $R$ has an energy level of conduction electrons with down-spin, while the $L$ has $N_\uparrow$ energy levels of conduction electrons with up-spin. 
%and $N_\downarrow$ ones with down-spin. 
%The interaction in this energy region, $V'$, is written by,
the interaction between the electron and the localized spin 
%$V'$, 
is written by,
%The expression is written by,
\begin{eqnarray}
\label{int2}
V=
J_\perp' \displaystyle{\sum_{n\uparrow=1}^{N_\uparrow'}} \left( 
{c}_{\mbox{\tiny $R_{\downarrow}$}}^\dag {c}_{\mbox{\tiny $L_{n\uparrow}$}} {S}_+  +
{c}_{\mbox{\tiny $L_{n\uparrow}$}}^\dag {c}_{\mbox{\tiny $R_{\downarrow}$}} {S}_-
\right), 
\end{eqnarray}
where the suffix $R_{\downarrow}$ ($L_{n\uparrow}$) 
denotes the level of conduction electrons with the down-spin of the $R$ 
(the $n$th level of conduction electrons with the up-spin of the $L$), and 
$J_\perp'$ is 
%Further, $J_\perp$ ($J_z$) is 
the transverse exchange integral 
between the localized spin and the electron in this energy region. 
% in this energy region, 
%respectively. Further, 
%and $N_\uparrow'$ is the number of levels of the $L$ in this energy region. 
%between the electron spin and the localized spin, 
%and $V_0$ is a coefficient for the spin independent transport. 
%left and right interfaces represent that 
%the interfacial layer with sp
%In Fig. \ref{sw_dev1}, 
%it is shown that 
%On the other hand, Fig. \ref{sw_dev2} represents that 
%When the voltage of the right electrode 
%(the voltage of the left electrode) 
When the bias voltage 
%(the voltage of the left electrode) 
is applied to the right electrode 
as shown in Fig. \ref{sw_dev1} (b), 
%so as to meet 
%the energy level of the $R$, 
%so that the highest occupied levels of the both electrodes 
%so that the Fermi level 
%of the $R$ and $L$ 
%becomes $E_{4,R}$ which satisfies $E_{3,L}$=$E_{4,R} - [E_{2,R} + e^2/(4\pi \epsilon r)]$, 
%respectively, 
%and the energy level of the left electrode is set to be $E_2$, 
the down-spin electrons are injected from the $R$ into the dot 
and the reversal from $S_z$=$S$ to $-S$ is achieved 
through the interaction of Eq. (\ref{int2}). 
%by injecting the down-spin electron from the $R$ into the $L$ 
%[see Fig. \ref{sw_dev1} (b)]. 
%Thus, we speculate that 
%%%in this sytem, 
%the reversible device is realized 
%by controlling the voltage of the elctrodes. 
%\end{enumerate}
%Note that the application of the voltage to the left electrode eliminates the transport in the region of $E < E_{2,L}$. 
%Note that the application of the voltage to the left electrode is applied to eliminate the transport in the region of $E < E_{2,L}$. 

%%%%%%%%% Quasi-Fermi Level - the energy level at which the occupation probability is equal to1/2 within a single energy band 
%% The Fermi level is the highest occupied level at. T = 0

%%%%which will play a central role in our analysis of the eproton spinf problem. It is called the ...... To conclude, we shall make some brief comments about the experimental. requirements necessary for these predictions to be tested[ 

%Finally, we discuss the experimental situation 
Finally, we discuss the experimental aspects 
%on the basis of 
from the viewpoint of the magnitude of current density. 
%On the basis of 
In the case of $S$=5, $N_\downarrow$=50, and $|J_\perp|$=0.001 eV, 
%[see Fig. \ref{s5}]. 
%[see Fig. \ref{s5}]. 
%As seen from the middle panel of Fig. \ref{s5}, 
%As seen from the middle panel of this figure, 
the required maximum current density for the spin reversal, 
$j_{\mbox{\tiny max}}$, 
is evaluated to be 
$j_{\mbox{\tiny max}}$=
$I_{\mbox{\tiny max}}/S_d $ $\simeq$ 4.2 $\times$ 10$^8$ A/cm$^2$, 
%with $I_{\mbox{\tiny max}}$ being the maximum current 
%and it is given by As seen from the middle panel of this figure, 
where 
the maximum current $I_{\mbox{\tiny max}}$ is given by 
$I_{\mbox{\tiny max}}$ = 
$(5.9 \times 10^{13}) \times (1.6 \times 10^{-19})$ 
$\simeq$ 9.4 $\times$ 10$^{-6}$ A 
as shown in the middle panel of Fig. \ref{s5}, 
with $n_e S_d l$=1 and $e $=$1.6 \times 10^{-19}$ C. 
%and also 
%$\Delta t$ = 10$^{-13}$, 
Furthermore, $S_d$ is set to be $S_d$=(1.5 $\times$ 10$^{-7}$)$^2$ cm$^2$, 
%%Here, 
%$\Delta t$ is a typical value obtained in this study [see Fig. @@], 
%the value of 
%This $S_d$ 
which is obtained by assuming that 
the molecule with $S$=5 has a cubic structure that is about 1.5 nm per side. 
%where 
This length is roughly estimated 
using examples from
a similar molecule, Mn$_{12}$ with $S$=10.~\cite{Sessoli} 
%Actually, the above mentioned $j_{\mbox{\tiny max}}$ is extremely large for 
%In practice, the above mentioned $j_{\mbox{\tiny max}}$ may be rather high 
In terms of sustainability, 
the abovementioned $j_{\mbox{\tiny max}}$ is high 
for typical ferromagnetic tunnel junctions, 
such as CoFeB/MgO/CoFeB junctions with 
critical current densities for the MRSI of 
7.8 $\times$ 10$^5$ A/cm$^2$ and 
1.2 $\times$ 10$^7$ A/cm$^2$.~\cite{Hayakawa1} 
%Actually, it may be difficult for the present experimental situation to apply such a large $j$ without the breaking of junctions.~\cite{Hayakawa1} 
%appars to be difficult for . 
On the other hand, it has been recently reported that 
some carbon nanotubes can individually carry currents 
with density exceeding $10^9$ A/cm$^2$.~\cite{CNT1, CNT2} 
%However, it has been recently reported that any carbon nanotubes exhibit $j>10^9$ A/cm$^2$.~\cite{CNT} 
We therefore anticipate that 
appropriate use of carbon nanotubes encapsulating 
magnetic molecules~\cite{CNT_mag} 
%the molecule with the quantum spin 
%are utilized in the present system, 
will lead to the achievement of the spin reversal. 
%%%%%%% Therefore, if carbon nanotubes encapsulating the magnetic molecule with the quantum spin are utilized in the present system, the quantum spin reversal may be realized. 
%We expect that the present device may be fabricated by utilizing carbon nanotubes encapsulating the magnetic molecule. 
%the present spin reversal can be done in 
%a system such as the carbon nanotube encapsulating the Mn$_12$. 
In addition, we strongly believe that 
the current density can be tuned by controlling 
the exchange integral and $N_\uparrow$ (or $N_\downarrow$), 
as found from Eq. (\ref{current}). 
Namely, the current density decreases 
%becomes small 
with reducing the magnitude of the exchange integral 
and $N_\uparrow$ (or $N_\downarrow$).

\section{Conclusion}
\label{conclusion}

We studied the localized quantum spin reversal 
%by the sequential injection of spins in the dot. 
%We considered 
for a case in which 
up-spin electrons were sequentially injected into the spin quantum dot. 
To describe the sequentially injected electrons, 
we first made assumptions about the sequential 
%processes 
injection of spins 
and then proposed a simple method 
%we proposed a simple method 
based on approximate solutions 
from the time-dependent Schr$\ddot{\rm o}$dinger equation. 
This method was applied to systems with the following $R$; that is, 
(a) the spin-polarized $R$ having 
energy levels of conduction electrons with only down-spin, 
and 
(b) the non-magnetic $R$. 
% having energy levels of conduction electrons with equal number of up and down-spins. 
The respective results are summarized below. 
%The respective results are as follows: 

\begin{enumerate}
\item[(a)]
The system exhibited a spin reversal. 
In particular, 
we derived the expression of the reversal time of the localized spin 
and explicitly showed the upper and lower limits of 
%the reversal time 
this time 
for systems with $S \ge 1$. 
%This expression is useful for the rough estimation of the minimum relaxation time of the localized spin to achieve the spin reversal. 
%For the system of $S$=5,
%We also obtained 
In addition, analytic expressions for 
%the expectation value of localized spin 
%and the velocity of electron. 
%We also obtain characteristic time dependences of 
the expectation value of the localized spin and 
the electrical current 
%, and the number of electrons in the $R$ and the right electrode 
were obtained as a function of time. 
%We also showed 
%They exhibited the characteristic time dependences. 
% of them were also shown. 
% elucidated
%of the expectation value of localized spin and the velocity of the electron. 
\item[(b)]
%It was also shown that 
The system of $S$=2 and $N_\uparrow$=$N_\downarrow$=20 
exhibited the spin reversal or non-reversal 
depending on the exchange integral. 
%The spin reversal occured 
%when 
%Moreover, 
In the spin non-reversal case, 
the second injected electron was trapped in the dot, 
%while 
although 
all injected electrons transferred from the $L$ to the $R$ 
for the spin reversal case. 
%the $m$th injected electron with $m \ne 1$ was trapped in the dot. 
%Note also that the qualitatively same behaviors were found for systems with the other $S$. 
\end{enumerate}

%We expect that the characteristic phenomena 
We expect that the above-described phenomena 
will be observed with the advancement of experimental techniques 
and will be utilized 
in spin electronics devices in the future.

\begin{acknowledgments}
This work has been supported by 
a Grant-in-Aid for Young Scientists (B) 
(No. 18710085) from 
the Japan Society for the Promotion of Science. 
\end{acknowledgments}

\section{Appendix: bias and gate voltages for the SET}
\label{Appendix}
%Using Eq. (\ref{}), 
%In the plane of the bias and gate voltages, 
The SET region in the plane of the bias and gate voltages 
for the model of Fig. \ref{model} is found. 
%For this circuit, 
We investigate the energy change $\Delta E$ 
which is obtained by subtracting the energy of the initial charge state 
from that of the final charge state. 
%changing the charged state. 
%In the case of $\Delta E < 0$, the change of state occurs. 
The change of state really occurs for $\Delta E < 0$.

\begin{figure}[ht]
\begin{center}
\includegraphics[width=0.9\linewidth]{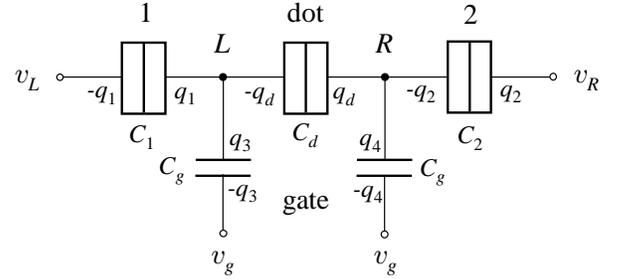}
\caption{
%The spin quantum dot. 
Equivalent circuit 
for ``electrode/$L$/spin quantum dot/$R$/electrode" junctions. 
The tunnel barrier between the left electrode 
and the $L$ (between the $R$ and the right electrode) is named 1 (2). 
%The symbol formed by two rectangles represents 
Each symbol composed of two rectangles represents 
both the condenser and the tunnel junctions 
which enable electron tunneling to take place. 
Each condenser of the gate 
%characterized by $C_g$ 
is assumed to have a thick barrier 
that suppresses the electron tunneling. 
%Here, 
%Here, $L$ ($R$) represents the left (right) quantum well. 
% The $L$ has an energy level of conduction electrons with up-spin, while the $R$ has $N_\uparrow$ energy levels of conduction electrons with up-spin and $N_\downarrow$ ones with down-spin. The shaded area in the electrodes represents the region occupied by electrons. A voltage is applied to the left electrode. 
}
\label{circuit}
\end{center}
\end{figure}

%in the plane of the bias voltage $v$ and gate voltage $v_g$. 
The equivalent circuit for this model 
is shown in Fig. \ref{circuit}. 
%Here, 
The bias voltage of the left electrode (the right electrode) 
is represented by $v_L$ ($v_R$), 
and the voltage of the gate electrode is $v_g$. 
The capacitances of junctions of 1, 2, the dot, and the gate are 
$C_1$, $C_2$, $C_d$, and $C_g$, respectively. 
The electric charge of the respective junctions is $q_i$~($i$=1-4, $d$). 
%$C_1$, $C_2$, $C_d$ and $C_g$ are capacitances for condensers of 1, 2, dot, and gate, respectively. 
%The capacitances of the junctions are C1and C2. The cluster is capacitively coupled to the gate, with capacitance CGC1 ,C2.
%where $q_i$~($i$=1-4, $d$) is the electric charge of the respective junctions [see Fig. \ref{circuit}]. 
%The SET region in the plane of the bias and gate voltages. 
%the region to bring about the SET. 

%In Fig. \ref{circuit}

When 
the initial charge at the $L$ ($R$) is represented by $q_L$ ($q_R$), 
the final charge at the $L$ ($R$) is $q_L'$ ($q_R'$), and 
the change of the charge number 
in the left electrode (right electrode) is $\Delta n_{Le}$ ($\Delta n_{Re}$), 
%is changed 
%We consider the change of state 
%from $(q_L, q_R)$ to $(q_L', q_R')$, 
%from $(q_L, q_R, n_{Le}, n_{Re})$ to $(q_L', q_R', n_{Le}', n_{Re}')$, 
%In the case of $\Delta E < 0$, the change of state really occurs. 
the energy change 
$\Delta E_{q_L, q_R, q_L', q_R', \Delta n_{Le}, \Delta n_{Re}}$ is given by, 
\begin{eqnarray}
\Delta E_{q_L, q_R, q_L', q_R', \Delta n_{Le}, \Delta n_{Re}} 
= \Delta U - W_1 -W_2 - W_3 - W_4, \nonumber \\
%\Delta E = \Delta U - W_1 -W_2 - W_{g_1} - W_{g_2},
\end{eqnarray}
where $\Delta U$ denotes the change of the electrostatic energy, 
and 
$W_1$ ($W_2$) represents the work done by 
the voltage of the left electrode (the right electrode) 
and $W_3$ ($W_4$) is that of the left gate electrode  
(the right gate electrode). 
%When the bias voltage of the left electrode (the right electrode) is written by $v_L$ ($v_R$) and that of the gate electrode is $v_g$, 
These expressions are written as follows: 
\begin{eqnarray}
&&\Delta U = U' - U, 
\\
&&U=\frac{C_1}{2}( \phi_L - v_L)^2  + \frac{C_2}{2} ( v_R - \phi_R )^2
+ \frac{C_d}{2} ( \phi_R - \phi_L)^2 \nonumber \\
&&\hspace*{.9cm}+ \frac{C_g}{2} (\phi_L - v_g)^2
+ \frac{C_g}{2} (\phi_R - v_g)^2, \\
&&U'=\frac{C_1}{2}( \phi_L' - v_L)^2  + \frac{C_2}{2} ( v_R - \phi_R' )^2
+ \frac{C_d}{2} ( \phi_R' - \phi_L')^2 \nonumber \\
&&\hspace*{.9cm}+ \frac{C_g}{2} (\phi_L' - v_g)^2
+ \frac{C_g}{2} (\phi_R' - v_g)^2, \\
\label{w_1}
&&W_1=-\Delta q_1 v_L - e\Delta n_{Le} v_L, \\
\label{w_2}
&&W_2= \Delta q_2 v_R + e\Delta n_{Re} v_R, \\
%&&W_1= \Delta q_1 v_R + \Delta n_1 e v_1, \\
%&&W_2=-\Delta q_2 v_2 - \Delta n_2 e v_2, \\
&&W_3=-\Delta q_3 v_g, \\
&&W_4=-\Delta q_4 v_g, 
\end{eqnarray}
with 
\begin{eqnarray}
%&&\Delta n_{Re}=n_{Re}' - n_{Re}, \\
%&&\Delta n_{Le}=n_{Le}' - n_{Le}, \\
&&\Delta q_1 
= C_1 ( \phi_L'-v_L) - C_1 (\phi_L - v_L), \\ 
&&\Delta q_2 
= C_2 ( v_R - \phi_R') - C_2 (v_R - \phi_R), \\ 
%= C_1 (\phi_A - \phi_A'), \\
%= C_2 (\phi_B' - \phi_B), \\
&&\Delta q_3 
= C_g (\phi_L' - v_g) - C_g ( \phi_L - v_g), \\
%= C_g (\phi_B' - \phi_B), \\
&&\Delta q_4 
= C_g (\phi_R' - v_g) - C_g ( \phi_R - v_g), \\ 
%= C_g (\phi_A' - \phi_A), \\
\label{phi_L}
&&\phi_L=\alpha q_R + \beta q_L + \gamma, \\
\label{phi_R}
&&\phi_R=\zeta q_R + \alpha q_L + \eta, \\
&&\phi_L'=\alpha q_R' + \beta q_L' + \gamma, \\
&&\phi_R'=\zeta q_R' + \alpha q_L' + \eta, \\
&&\alpha=\frac{C_d}{X}, 
\\
&&\beta=\frac{1}{X}(C_2 + C_d + C_g), 
\\
&&\gamma=\frac{1}{X} \left[ C_d (C_2 v_R + C_g v_g)\right. \nonumber \\
&&\hspace*{.9cm}\left. + (C_2 + C_d + C_g)(C_1 v_L + C_g v_g)\right], 
\\
&&\zeta=\frac{1}{X}(C_1 + C_d + C_g), \\
&&\eta=\frac{1}{X} \left[ C_d (C_1 v_L + C_g v_g) \right. \nonumber \\
&&\hspace*{.9cm}\left. + (C_1 + C_d + C_g)(C_2 v_R + C_g v_g) \right], \\
&&X=(C_1 + C_d + C_g)(C_2 + C_d + C_g) - C_d^2, 
\end{eqnarray}
%with $\Delta n_1$=$n_1' - n_1$ and $\Delta n_2$=$n_2' - n_2$. 
where 
$e$ ($>$0) is the electric charge. 
%, and 
%$U$ and $U'$ are the Coulomb potential energy of the initial state (the final state), and 
%$C_1$, $C_2$, $C_d$ and $C_g$ are capacitances for condensers of 1, 2, dot, and gate, respectively. 
%The changes of the charge number, $\Delta n_{Le}$ and $\Delta n_{Re}$, 
The works, $- e\Delta n_{Le} v_L$ of Eq. (\ref{w_1}) 
and $e\Delta n_{Re} v_R$ of Eq. (\ref{w_2})
represent that 
the electrodes are provided with 
%the electrodes provide with 
the changed portion of the charge due to the tunnel electron. 
%In addition, we note that 
Further, 
$\phi_L$ ($\phi_R$) is the electrostatic potential of the $L$ ($R$), 
and Eqs. (\ref{phi_L}) and (\ref{phi_R}) are 
obtained from the following equations, 
\begin{eqnarray}
&&q_L = -q_d + q_1 + q_3, \\
&&q_R = -q_2 + q_d + q_4, 
\end{eqnarray}
with 
$q_1$=$C_1 (\phi_L - v_L)$, 
$q_2$=$C_2 (v_R - \phi_R)$, 
$q_3$=$C_g (\phi_L - v_g)$, 
$q_4$=$C_g (\phi_R - v_g)$, 
$q_d$=$C_d (\phi_R - \phi_L)$. 
%where $q_i$~($i$=1-4, $d$) is the electric charge of the respective junctions [see Fig. \ref{circuit}]. 
%the Coulomb island. 
%We consider the change of state from $(q_A, q_B)$=$-e(0,1)$ to $(q_A', q_B')$=$-e(0,1)$. When $\Delta E < 0$, the change of state occurs. The other changes are confirmed in similar way. 
%Furthermore, 
% is satisfied, 
%the bias voltage $E_{\rm vol}$ and the gate voltage $E_{\rm gate}$. 
%For simplicity, 
%By setting $v_L$ ($v_R$) to be $-v/2$ ($v/2$), 
The bias voltage $v_L$ ($v_R$) is now set to be $-v/2$ ($v/2$). 
%we obtain 
When $\Delta E_{q_L, q_R, q_L', q_R', \Delta n_{Le}, \Delta n_{Re}}$=0, 
the relation between $v$ and $v_g$ 
%for the case of 
is given by, 
%that is, 
%the relation is given as follows: 
%an expression of lines in Fig. @@@. 
%the relation between the bias voltage $E_{\rm vol}$ 
%and the gate voltage $E_{\rm gate}$. 
\begin{eqnarray}
\label{vol_gate}
v=\frac{-\left[ \lambda (d_1 - d_3) + \nu (d_2 + d_3 ) \right] 
v_g - d_4 }{\kappa (d_1 -d_3) + \mu (d_2 + d_3) 
- \frac{1}{2}( \Delta n_{Re} + \Delta n_{Le})e }, 
\end{eqnarray}
with
\begin{eqnarray}
&&d_1=(C_1 + C_g) \left[ \alpha ( q_R' - q_R) + \beta ( q_L' - q_L) \right], \\
&&d_2=(C_2 + C_g) \left[ \zeta (q_R' - q_R) + \alpha ( q_L' - q_L) \right], \\
&&d_3=C_d \left[ (\zeta - \alpha) (q_R' - q_R) + ( \alpha - \beta) ( q_L' - q_L) \right], \nonumber \\\\
&&d_4=\frac{C_2 + C_g}{2} \left[ - ( \zeta q_R + \alpha q_L )^2 + ( \zeta q_R' + \alpha q_L')^2 \right] \nonumber \\
&&\hspace*{.9cm}+ \frac{C_1+ C_g}{2} \left[ (\alpha q_R' + \beta q_L')^2 - ( \alpha q_R + \beta q_L)^2 \right] \nonumber \\
&&\hspace*{.9cm}+ \frac{C_d}{2} \left\{ \left[ (\zeta - \alpha) q_R' + (\alpha - \beta) q_L' \right]^2 \right. \nonumber \\
&&\hspace*{.9cm} \left. - \left[ (\zeta -\alpha) q_R + (\alpha - \beta) q_L \right]^2 \right\}, \\
&&\kappa=\frac{1}{2X} (C_2 C_d - C_1 C_2 - C_1 C_d - C_1 C_g), 
\\
&&\lambda=\frac{1}{X} (C_2 C_g + 2 C_d C_g + C_g^2 ), 
\\
&&\mu=\frac{1}{2X} (C_1 C_2 + C_2 C_d + C_2 C_g - C_1 C_d), 
\\
&&\nu=\frac{1}{X} (C_1 C_g + 2 C_d C_g + C_g^2 ). 
\end{eqnarray}

\begin{figure}[ht]
\begin{center}
\includegraphics[width=0.8\linewidth]{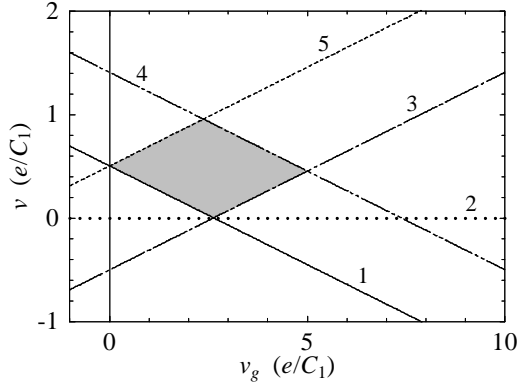}
\caption{
%The spin quantum dot. 
SET region in the bias ($v$) and gate ($v_g$) voltage plane 
%$v$-$v_g$ plane 
for the equivalent circuit 
with $v_L$=$-v/2$, $v_R$=$v/2$, 
$C_1$=$C_2$, $C_d$=10$C_1$, and $C_g$=$0.1C_1$. 
%of Fig. \ref{circuit}. 
%Here, $C_1$=$C_2$, $C_d$=10$C_1$, and $C_g$=$0.1C_1$ is set. 
This region is shown by the shaded area 
%the specific $v$ and $v_g$, 
which satisfies 
$\Delta E_{0, 0, -e, 0, 1, 0} < 0$, 
$\Delta E_{-e, 0, 0, -e, 0, 0} < 0$, 
$\Delta E_{0, -e, 0, 0, 0, 1} < 0$, 
$\Delta E_{0, -e, -e, -e, 1, 0} > 0$, 
and $\Delta E_{0, 0, 0, +e, 0, 1} > 0$. 
The meaning of each line is as follows: 
1: $\Delta E_{0, 0, -e, 0, 1, 0}$=0, 
2: $\Delta E_{-e, 0, 0, -e, 0, 0}$=0, 
3: $\Delta E_{0, -e, 0, 0, 0, 1}$=0, 
4: $\Delta E_{0, -e, -e, -e, 1, 0}$=0, 
5: $\Delta E_{0, 0, 0, +e, 0, 1}$=0, 
where the expression is given by Eq. (\ref{vol_gate}). 
}
\label{SET_region}
\end{center}
\end{figure}

%The SET is realized when $\Delta E < 0$, $\Delta E < 0$, $\Delta E < 0$, $\Delta E > 0$ are satisfied. 
This model exhibits the SET 
under the specific $v$ and $v_g$, 
which satisfy 
$\Delta E_{0, 0, -e, 0, 1, 0} < 0$, 
$\Delta E_{-e, 0, 0, -e, 0, 0} < 0$, 
$\Delta E_{0, -e, 0, 0, 0, 1} < 0$, 
$\Delta E_{0, -e, -e, -e, 1, 0} > 0$, 
and $\Delta E_{0, 0, 0, +e, 0, 1} > 0$. 
%, in which the electrons are injected from the left electrode into the dot one by one
%$E_{\rm vol}$-$E_{\rm gate}$ 
%the region satisfying $\Delta E < 0$, $\Delta E < 0$, $\Delta E < 0$, $\Delta E > 0$. 
%For the case of 
%For $C_1:C_2:C_d:C_g$=$1:1:1:0.1$, 
%the $E_{\rm vol}$-$E_{\rm gate}$ region 
%to bring about the SET 
In Fig. \ref{SET_region}, 
the SET region for a system with 
$C_1$=$C_2$, $C_d$=10$C_1$, and $C_g$=$0.1C_1$ 
is shown by the shaded area.

%The shaded area in Fig. @@@ shows 

We furthermore describe the present transport property; that is, 
%Furthermore, under such voltages, 
%we consider that 
when the probability density of the injected electron becomes 0 at the $L$ 
and 1 at the $R$, 
the electron moves into the right electrode, 
%and the subsequent electron is injected into the $L$, 
%We here assume that 
while the electron cannot go to the right electrode 
if the probability density is finite for both the $L$ and the $R$. 
%the electron cannot go out of the $L$-dot-$R$ system. 
This property is based on the following two assumptions: 
First, there is no direct transfer integral 
between the $L$ and the right electrode. 
%or between the $R$ and the left electrode, 
%although that between the $R$ and the right electrode is present. 
%is considered. 
%or between the $R$ and the left electrode
%negligile small compared to that between the $R$ and the right electrode or that between the $L$ and the left electrode. 
Second, 
%It is therefore assumed that 
%It is assumed that 
the wave function extended over 
%the wave function does not expand 
both 
%between 
the $R$ and the right electrode can be ignored, 
%is not taken into account, 
%while 
although that extended over both the $L$ and the $R$ is taken into account, 
%because 
%provided that 
under the condition that 
the coupling between the $R$ and the right electrode 
%or that between the right electrode and $R$ are 
is much smaller than 
that between the $L$ and the $R$, i.e. Eq. (\ref{int}).

%wave function is extended over the two dots 
%In metallic nanotubes, the wave function is extended in the whole nanotube because of the absence of
%wave function is extended into a sufficiently large region of space


\begin{thebibliography}{10}
\bibitem{comment}
As for the spin-polarized $R$, 
by using Eq. (\ref{delta_sz}), 
%By using Eq. (\ref{omega_Sz}), 
%we find that 
%the relation of 
$|\Delta_{R_{n\downarrow},S_z}/ F_{S_z}| \ll 1$ 
%and $|\Delta_{R_{n\uparrow},S_z}/ G_{S_z}| \ll 1$ 
is found to correspond to 
$\Delta t^{(m)} \ll \hbar/| \Delta_{R_{n\downarrow},S_z}|$ 
[see Sec. \ref{assump}(iv)]. 
%with $\Delta E \equiv | \Delta_{R_{n\downarrow},S_z}|$. 
%Furthermore, 
For the non-magnetic $R$, 
$|\Delta_{R_{n\downarrow},S_z}/ F_{S_z}| \ll 1$ 
and $|\Delta_{R_{n\uparrow},S_z}/ G_{S_z}| \ll 1$ 
are related to $\Delta t^{(m)} \ll \hbar/| \Delta_{R_{n\downarrow},S_z}|$ 
by taking into account $\Delta t^{(m)}$ in Table \ref{tab1}.
\bibitem{comment1}
In the case of spin reversal of non-magnetic $R$, 
$t$ corresponds to $\Delta t^{(m)}$ in Table 1 at maximum, 
which is $3\pi/2$ times as large as $1/\Omega_{S_z}$. 
In the case of spin-polarized $R$, 
$t$ is  Eq. (\ref{delta_sz}) at maximum, 
which is $\pi/2$ times as large as $1/\omega_{S_z}$.
\end{thebibliography}
\end{document}